\documentclass[usenatbib]{mn2e}
\usepackage{graphicx}
\usepackage[dvips]{color}
\usepackage{amssymb}

\newcommand{\ale}{\ \raisebox{-.3ex}{$\stackrel{<}{\scriptstyle \sim}$}\ }
\newcommand{\age}{\ \raisebox{-.3ex}{$\stackrel{>}{\scriptstyle \sim}$}\ }

\begin{document}

\title[Particle evolution in protoplanetary discs]{Global variation of the dust-to-gas ratio in\\ evolving protoplanetary discs}
\author[Hughes \& Armitage]{Anna L.H. Hughes$^{1,2}$ \& Philip J. Armitage$^{1,2}$ \\
$^1$JILA, University of Colorado, Boulder \& NIST, 440 UCB, CO 80309-0440, USA \\
$^2$Department of Astrophysical and Planetary Sciences, University of Colorado, Boulder, USA \\
\tt e-mail: Anna.Haugsjaa@Colorado.EDU (ALHH), pja@jilau1.colorado.edu (PJA)}

\maketitle
  
\begin{abstract}
Recent theories suggest planetesimal formation via streaming and/or gravitational instabilities may be triggered by localized enhancements in the dust-to-gas ratio, and one hypothesis is that sufficient enhancements may be produced in the pile-up of small solid particles inspiralling under aerodynamic drag from the large mass reservoir in the outer disc. Studies of particle pile-up in static gas discs have provided partial support for this hypothesis. Here, we study the radial and temporal evolution of the dust-to-gas ratio in turbulent discs, that evolve under the action of viscosity and photoevaporation. We find that particle pile-ups do not generically occur  within evolving discs, particularly if the introduction of large grains is restricted to the inner, dense regions of a disc. Instead, radial drift results in depletion of solids from the outer disc, while the inner disc maintains a dust-to-gas ratio that is within a factor of $\sim 2$ of the initial value. We attribute this result to the short time-scales for turbulent diffusion and radial advection (with the mean gas flow) in the inner disc. We show that the qualitative evolution of the dust-to-gas ratio depends only weakly upon the parameters of the disc model (the disc mass, size, viscosity, and value of the Schmidt number), and discuss the implications for planetesimal formation via collective instabilities.
 Our results suggest that in discs where there is a significant level of midplane turbulence and accretion, planetesimal formation would need to be possible in the absence of large-scale enhancements.   
 Instead, trapping and concentration of particles within local turbulent structures may be required as a first stage of planetesimal formation.

\end{abstract}

 
\section{Introduction}
\label{Introduction}

Uncovering the pathway by which sub-micron dust grains grow into gravitationally bound bodies suitable for building planets is a long-standing problem.   Pairwise collisions between dust particles in protoplanetary discs are thought to drive  rapid growth up to $\sim$mm sizes \citep{DullemondDominik2005}. Beyond this size scale, continued growth up to the km-scale of
planetesimals  is potentially frustrated by two distinct physical barriers. First, the measured material properties of  solid aggregates, which favor sticking upon collision for small sizes, switch to favour bouncing  \citep{Guttler2010,Zsometal2010}, or collisional fragmentation due to high, turbulence-driven velocity dispersions \citep[e.g.][]{BirnstielDullemondBrauer2010} for $s_d < {\rm m}$. Even if these barriers could be surmounted, however, direct collisional growth faces a second obstacle from headwind drag \citep{Weidenschilling1977a}, which becomes extremely strong for particles in the cm to m-size range. The loss of substantial quantities of solids into the star is not in itself deleterious -- indeed the relatively small solid mass in the Minimum Mass Solar Nebula \citep{Weidenschilling1977b} indicates loss probably occurred -- but the residence time of critically coupled particles is so short that it is hard to see how catastrophic depletion could be averted if bodies grow steadily through m-scales. 

These considerations have led to renewed interest in collective mechanisms for planetesimal formation, which appeal (typically) to gravitational instability in a dense particle layer \citep{Safronov72,Goldreich73} to effect a dynamical time scale jump from mm or cm-sizes to km-scales \citep[for a review see][]{ChiangYoudin2010}. Several precursor mechanisms have been proposed that might act to create conditions locally favourable for gravitational instability. They include vertical settling \citep{YoudinShu2002}, turbulent concentration \citep{CuzziHoganShariff2008,Johansenetal2007}, and streaming instabilities \citep{JohansenYoudin2007,JohansenYoudinMacLow2009,BaiStone2010b}. There is some evidence that these mechanisms could form surprisingly large first-generation planetesimals ($\age 100 \ {\rm km}$), of a size consistent with that derived --- by combining the measured size distribution with the results of collisional-grinding simulations ---  for the primordial asteroid belt  \citep{Morbidellietal2009b}.

 Which, if any, of these mechanisms is responsible for planetesimal formation is hard to say, since their feasibility depends in part on the uncertain properties of the intrinsic turbulence within the disc \citep{Armitage2011}. Some theoretical constraints are nonetheless possible, since most of the formation-via-collapse models require local enhancement of the height-averaged dust-to-gas ratio (the local disc metallicity) above solar values in order to exceed the threshold for gravitational instability
\footnote{The \cite{CuzziHoganShariff2008} model of planetesimal formation 
may be an exception as it considers only height-local dust/gas enhancement, possibly achievable by vertical settling alone.}.  
 The magnitude of the required enhancement varies both with the model, and with the properties of the gas disc, but typical values quoted vary between roughly 2 and 10 times standard metallicity \citep{YoudinShu2002,JohansenYoudinMacLow2009,BaiStone2010b}. Either a
local build-up of solids or a local depletion of gas \citep{ThroopBally2005,AlexanderArmitage2007} would be equally effective. 

Further clues come from observations.  To match Solar System constraints, conditions for planetesimal formation must extend over a significant radial range (because planetesimals formed with substantially distinct compositions) and persist for a large fraction of the disc lifetime. The evidence for an extended era of planetesimal formation comes from age dating of primitive meteorites, and in particular from the finding that the chondrules within those meteorites differ in age by as much as 2.5~Myr \citep{Amelinetal2002}.  Furthermore, recent age-dating of iron-core meteorites suggests the era of planetesimal formation extended nearly as far back as the beginning of Solar-system history, as defined by the formation of calcium-aluminium-rich inclusions (CAIs) \citep{Kleineetal2005,Bottkeetal2006}.

The replenishment of cold dust in debris discs \citep{Wyatt2008} implies that planetesimal populations around other stars are not solely confined to the inner disc, but there is no observational requirement that planetesimal formation spans the entire range of radii where gas was once present. Given this, one obvious way to create an enhancement of the dust-to-gas ratio at small to intermediate disc radii is to appeal to the large reservoir of mass typically present further out. If small particles spiralling inward pile-up as they migrate radially, then the ``problem" of headwind drag becomes instead a necessary ingredient for planetesimal formation. In a static (inviscid) disc model, headwind drag acting on an initially uniform dust-to-gas ratio does in fact result in just such a pile-up, with peak values of the enhancement exceeding those needed to produce gravitational collapse via vertical sedimentation \citep{YoudinShu2002,Youdin2004}. Although not the focus of the original work, radial pile-ups would also result in highly favorable conditions for triggering streaming instabilities. 

Inviscid-disc models, which assume that the gas has neither a net radial flow nor local turbulent motions, include only the minimal set of potential aerodynamic effects that can modify the radial particle distribution. Such models also provide a poor representation of the global disc environment, which is expected \citep[and observed, e.g.][]{Hartmann1998} to change substantially over the time period of interest for planetesimal formation.  
Our goal in this paper is to quantify the variation in the dust-to-gas ratio if the gas disc, instead of being static, evolves with time under the action of local turbulent transport of angular momentum and photoevaporation. Headwind drag will still occur in an evolving disc, but the propensity of particles to pile-up will additionally be affected by several new processes,

\begin{itemize}
\item
The global evolution of the gas disc, which thins with time and whose outer edge must expand radially to conserve angular momentum as gas is accreted on to the star.
\item
Radial advection with the flow of disc gas, 
which may counter the slow-down of drift that leads to pileup in the inner disc.  
\item
Turbulent diffusion.
\item 
Disc-clearing photoevaporation.
\end{itemize}
The first three of these processes \citep[also explored in, e.g.,][]{StepinskiValageas1996} are ultimately byproducts of the same turbulence within the disc, so the inclusion of evolution introduces fewer independent parameters into a calculation of particle transport than one might at first fear. Indeed, the primary conclusion of our work is  that a range of disc models all produce roughly the same 
result: as 
the disc evolves, the outer disc becomes depleted of solids, but the lost material does not pile-up significantly in the inner
disc. Instead, the inner disc maintains roughly its primordial dust-to-gas ratio as the disc drains away.  
As time progresses, the thinning gas supports successively smaller particle sizes across the range of Solar
System radii.  

The plan of this paper is as follows.  In \S\ref{methods} we outline the one dimensional disc evolution and particle transport models that we employ. These models immediately specify the radial dependence of several critical time scales relevant to particle transport, and in \S\ref{Consequences} we discuss the elementary consequences of those time scales for dust evolution. In \S\ref{Results}, we show the dust distributions resulting from our simulations, which we compare in \S\ref{Results_VEnhancements} to the dust-to-gas enhancements thought necessary to lead to precipitation (formation via vertical sedimentation as in \cite{YoudinShu2002}) of large bodies, and, briefly in \S\ref{CompareObservations}, to observations of dust in discs around other stars. We discuss the implications of our results for planetesimal formation models in \S\ref{DiscussionConclusions}.

\section[Methods]{Methods}
\label{methods}

\subsection{Gas disc evolution}
\label{ModelDisc}

Our simulations assume a 1D, viscously evolving disc model, whose $t=0$ surface density profile is $\Sigma_g \propto R^{-1}$ at small radii, with a characteristic exponential fall-off from 
$R=R_d$,
\begin{eqnarray}
   \Sigma_{g}\left(R,t=0\right)
&=&
   \frac{M_{D,0} \,k_\mathrm{disc}}{\pi R}
   \left(1-\sqrt{\frac{R_\mathrm{in}}{R}}\right)
   \exp\left(-\frac{R}{R_d}\right) \,,
\nonumber \\
   k_\mathrm{disc}
&=&
   \frac{1}{2 R_d}
   \left( 1 - \sqrt{\frac{\pi R_\mathrm{in}}{R_d}} \right)^{-1}, 
\label{EDisc_sigma0}
\end{eqnarray}
where $\Sigma_g$ is the surface density of the disc gas, $M_{D,0}$ is the initial disc mass, $R$ is radial distance from the central star, and $R_\mathrm{in}=0.1$~AU is the inner boundary of the disc-model grid.  In the fiducial model, $M_{D,0}=0.03 M_\odot$ and $R_d=20$~AU, but simulations are also run in discs  with $M_{D,0}=0.01 M_\odot$ and $0.09 M_\odot$ or $R_d=5$~AU and 40~AU.

Viscous evolution of the disc follows the thin-disc model of disc evolution \citep{Pringle1981} on a 1D grid of 600 points spaced logarithmically in $R$ from 0.1 to 15,000~AU.  We use the \cite{ShakuraSunyaev1973} alpha-parameterization for the disc viscosity,
\begin{equation}
   \nu\left(R\right)
=
   \alpha c_s H_g
=
   \alpha \Omega_\mathrm{K} H_g^2 
   \,,
\label{Ealpha_prescription}   
\end{equation}
where $\nu$ is the viscosity, $c_s$ is the local gas sound speed, $\Omega_\mathrm{K}$ is the local Keplerian angular velocity, and $H_g = \Omega_\mathrm{K}^{-1}\sqrt{k_B T / \mu m_H}$, is the vertically-isothermal disc-gas scale-height.  $k_B$ is Boltzman's constant, and $\mu m_H$ is the average mass of a gas particle, set equal to 2.34 hydrogen atoms.  In the fiducial model, $\alpha=10^{-2}$, but simulations are also run in discs using $\alpha=5\times10^{-3}$ and $2\times10^{-2}$.  The disc-evolution model also includes photoevaporation at the end of the disc lifetime due to $10^{42}$ photons s$^{-1}$ of EUV radiation, as in \cite{HughesArmitage2010}, Equation (1).

We assume that the viscosity in Equation~(\ref{Ealpha_prescription}) derives from a local turbulent stress \citep{Balbus1999}. In this case, both the evolution of the gas disc profile, and the magnitude of the particle diffusivity, $D_p$, share a common scaling with the midplane disc temperature, $D_p \propto \nu \propto T$. Due to gas infall and accretional heating, disc temperatures at early times are much hotter than those of later discs \citep{Cassen2001}. To evaluate the evolution of the midplane temperature, we assume that local heating by viscous dissipation and stellar irradiation is balanced by radiative cooling. We compute these terms using standard methods (detailed in Appendix~\ref{AX_DiscTemp}), assuming a constant dust-to-gas ratio (i.e., ignoring coupling to the evolving dust distribution). For computational economy, when running our dust-transport simulations, we model the gas-disc evolution using a fit to the computed energy-balanced temperature evolution.  At a given time,
\begin{equation}
   T\left(R\right)
=
   \mathrm{max}\left[
   T_\mathrm{cloud} \,,\,
   T_\mathrm{AU}\left(t\right) \left(\frac{R}{1\mathrm{AU}}\right)^{q_T\left(t\right)}
   \right] \,,
\label{EDisc_Temp}
\end{equation}
where $T_\mathrm{cloud}=10$~K is the background temperature of the star-forming cloud, and $T_\mathrm{AU}$ and $q_T$ follow an exponential decay in time from an initially hotter, steeper distribution ($T_\mathrm{AU} \approx 500$~K, $q_T \approx -0.6$) to a cooler, shallower distribution appropriate for a passively-heated, flared disc ($T_\mathrm{AU} \approx 280$~K, $q_T \approx -0.5$). To compute the stellar irradiation, which dominates at late times and large radii, we assume a stellar luminosity of $L_\star=5 L_\mathrm{solar}$ and a disc-starlight intersection angle set to $\phi=0.05$ everywhere.  We have included the time-evolving nature of the disc-temperature distribution for completeness in considering the fundamental differences between an evolving versus a static disc-mass distribution.  However, at the time-scales / masses and accretion rates considered in our models, disc surface-density evolution is nearly identical using either a static or evolving temperature distribution.
We have also verified that using a fit to the evolving temperature distribution, rather than calculating it in each dust-transport simulation at every $(R,t)$, does not affect our dust-distribution results.

The disc-gas radial velocity, $v_{r,g}$, is calculated using the 1D accretional velocity derived due to the viscous evolution of the disc surface-density profile.
\begin{equation}
   v_{r,g}
=
   -\frac{3}{\Sigma_g R^{1/2}}
   \frac{\partial}{\partial R}\left(\nu \Sigma_g R^{1/2}\right)
   \,.
\end{equation}

This accretion-flow velocity is predominantly inward but includes outward flow at the outer edge of the expanding disc.  
The inclusion of the accretion velocity of the gas marks one of the primary differences between our evolving-disc model and the inviscid-disc model of \cite{YoudinShu2002}. It is important because, for a typical gas surface density profile ($\Sigma \propto R^{-1}$), $\vert v_{r,g} \vert$ does not decrease toward smaller disc radii, unlike the headwind-drag velocity.  
We note, however, that it is possible for discs to evolve without there being a significant accretion flow at the disc midplane. This situation occurs in dead-zone models \citep{Gammie1996,Armitage2011}, in which vertical variation in the efficiency of angular momentum transport results in gas accretion that occurs primarily along the disc surface.  Qualitatively, we expect that particle transport within dead-zone disc models would differ substantially depending on whether
the scale height of the particle disc was large enough for the particles to experience the (rapid) surface layer flow. However, more detailed study of these models is left to others.  

The gas azimuthal velocity, $v_{\phi,g}$, effecting headwind-drag on the simulation grains, is calculated at the disc midplane assuming pressure balance within the vertically-isothermal disc model \citep{TakeuchiLin2002}.  For the numerical calculation, we use
\begin{eqnarray}
   \Omega_g^2
&=&
   \Omega_\mathrm{K}^2
   \left(1 + \frac{H_g^2}{R^2}\left[q_\Sigma + \frac{1}{2}q_T - \frac{3}{2}\right]\right)
   \,,
\nonumber \\
   v_{\phi,g}
&=&
   R \Omega_g
   \,,
\end{eqnarray}
where $\Omega_g$ is the gas angular velocity, $q_T$ is the power-law index of the temperature distribution used in Equation~(\ref{EDisc_Temp}), and $q_\Sigma$ is the locally-calculated power-law index of the disc surface-density profile.

     \subsection[Particle Transport Setup]{Particle Transport Setup}
     \label{ParticleTransport}

          \subsubsection[Particle Motions]{Particle Motions}
          \label{ParticleTransport_Motion}

We use the particle-transport code outlined in \cite{HughesArmitage2010} to evolve the dust distribution, which is modeled as an ensemble of particles. Each particle is subject to aerodynamic advection and random-walk turbulent diffusion within the disc.  Particle trajectories are integrated following
\begin{equation}
   r_d\left(t+\Delta t\right)
=
   r_d\left(t\right) + \left(v_{srd}\pm v_\mathrm{turb}\right) \Delta t
   \,,
\label{ETransport_transport}
\end{equation}
where $r_d$ is a particle's radial position, $\Delta t$ the time-step, $v_{srd}$ is a particle's radial velocity due to gas-drag and advection within the mean gas flow, and $v_\mathrm{turb}$ is the mean velocity for effecting random-walk turbulent diffusion of the particle ensemble. The random walk is asymmetric, with the weighting between the inward and outward steps calculated from the local disc conditions so as to recover the evolution of a fluid model \citep{HughesArmitage2010}.

Gas drag is calculated assuming that particles experience Epstein drag appropriate to their size.  The Epstein drag regime is appropriate for the vast majority of the disc's radial and temporal extent, though the largest grains considered (cm-sized) technically fall within the Stokes-drag regime in the innermost disc ($\ale 1$~AU) at very early times.  The radial velocity of a grain is calculated at each radial grid-point by iteratively solving for the steady velocity of particles that have attained terminal velocity 
\citep{TakeuchiLin2002,HughesArmitage2010}, 
\begin{eqnarray}
   v_{srd}
&=&
   v_{r,g} +
   \frac{3 \left(v_{s\phi d}^2 - v_\mathrm{K}^2\right)}
        {C_R R \rho_g v_\mathrm{therm}}
   \,,
\label{ETransport_svrda}
\\
   v_{srd}
&=&
   - \frac{1}{3} C_R R \rho_g v_\mathrm{therm}
   \frac{\left(v_{s\phi d}-v_{\phi,g}\right)}
        {\left(v_{s\phi d}-v_\mathrm{K}/2\right)}
   \,.
\label{ETransport_svrdb}
\end{eqnarray}
Here $v_{s\phi d}$ is the dust azimuthal velocity, $v_\mathrm{K}$ is the local Kepler velocity, $C_R$ is the dust-grain surface-area-to-mass ratio, $\rho_g$ is the local gas density (set equal to the value at the midplane in a vertically isothermal disc), and $v_\mathrm{therm}$ is the gas thermal velocity.  We assume spherical grains of internal density 1 g cm$^{-3}$, and vary the grain radius, $s_d$.  We consider grain sizes of $2 s_d=0.2~\mu$m -- 2~cm, which is meant to represent the general range of grain sizes produced by coagulative grain-growth processes.  Above some critical size, laboratory experiments and coagulation models show that bouncing and collisional fragmentation represent barriers to further growth \citep{BrauerDullemondHenning2008,Zsometal2010}. The precise value of the critical size depends upon the gas disc conditions, and is not precisely known.   We assume 2~cm as the upper-limit grain size for tracking a global dust distribution.  Except in the case of instantaneous grain growth (discussed in \S~\ref{Results_VaryModel}), our simulation results are qualitatively identical for maximum assumed grain sizes as small as 0.2~mm.

The local velocity for turbulent diffusion is, 
\begin{equation}
   v_\mathrm{turb}
=
   \sqrt{2 D_p / \Delta t}
   \,,
\label{ETransport_vturb}
\end{equation}
where $D_p$ is the local diffusivity of the particle ensemble. We use the
\cite{YoudinLithwick2007} scaling for the particle radial diffusivity,
\begin{equation}
   D_p
=
   D_g
   \frac{1+4 \tau_s^2}
        {\left(1+\tau_s^2\right)^2}
   \,,
\label{ETransport_Dp}
\end{equation}
where $D_g=\nu/Sc$ is the gas-particle diffusivity, and $\tau_s=3\Omega_\mathrm{K} / \left(C_R \rho_g v_\mathrm{therm}\right)$ is the normalized gas-drag stopping time.  In general we assume a Schmidt number of $Sc=\nu/D_g=1$, but also run simulations in the fiducial-disc model using $Sc=1/2$ and 2.  Equation~(\ref{ETransport_Dp}) has been numerically validated by \cite{CarballidoBaiCuzzi2011}.  However, note that for $\tau_s \age 1$, it deviates from the earlier common expression given by \cite{CuzziDobrovolskisChampney1993}: $D_{p,CDC}=D_g / \left(1+St\right)$, where $St=\tau_s/\tau_e$, and $\tau_e$ is a normalized eddy turnover timescale.  We have run a comparison simulation of the evolving dust distribution within the fiducial disc model using this other dust-diffusivity parameterization (with commonly used $\tau_e=1$) and have found that the results are qualitatively identical to those using the parameterization in Equation~(\ref{ETransport_Dp}).  As discussed in \S\ref{Consequences}, headwind-drag generally confines grains to regions of the disc with $\tau_s < 1$.  

          \subsubsection[Particle Distributions]{Initial Particle Distribution and Dust Mass Allocation}
	  \label{ParticleTransport_Distribution}

We assume that, at $t=0$, the entire mass of disc solids is in small particles (0.2~$\mu$m) whose radial 
distribution matches that of the disc gas. Since coagulation occurs rapidly for such small-size particles, 
to a first approximation we expect that the grain-size distribution will rapidly approach an equilibrium 
dictated by collision/fragmentation processes (e.g., \cite{BirnstielOrmelDullemond2011}).  In more detail, however, the outer regions of a disc will experience a time-delay in the production of large grains relative to the inner, denser regions.  This should translate into a delay before which rapid inspiral of large grains can occur from the outer disc.  

The grain trajectories calculated in our simulations are non-coagulative, considering a single grain-size for the full lifetime of a given run.  In order 
to 
account for the delay in the appearance of larger grain sizes, we employ a pair of simplified grain-growth models (described in Appendix~\ref{AX_GrainGrowth}) to define a radially dependent timescale for the appearance of larger grains.  Larger grains are then initialized at time $t_0\left(R\right) > 0$ within ten radial zones for which grain-growth time scales (at characteristic radii) have been calculated.  
This initialization of all grains larger than 0.2~$\mu$m diameter occurs only once per zone, explicitly neglecting the collisional or coagulative regeneration and resupply of any grain-size population that may be removed due to transport effects.  However, as discussed in \S\ref{Results}, limiting the appearance of large grains (especially in the inner disc where they would be most likely to regenerate) has the primary effect of retaining that dust mass against loss inward onto the parent star. This simplification therefore results in the highest dust-to-gas ratio in the inner disc achievable with our disc-evolution and dust-transport models.  

Our simplified grain-growth models consider coagulation via differential settling (in the rapid-upper-limit case in the absence of turbulence) and random motions.  We do not consider growth via differential-radial drift, which, e.g., \cite{StepinskiValageas1996}, found to produce more rapid growth time-scales, nor directly via turbulent velocity dispersions.  Our assumed dust distributions, therefore, may represent a lower-limit on the rapidity and radial extent of large-grain initialization.  In \S\ref{Results_VaryModel}, we consider the upper limit cases 
where large-grain formation is a) instantaneous, and b) independent of disc location (not confined to the inner regions of the disc) 
in order to 
define the boundaries of 
the behavior of our evolving dust/gas distributions within the grain-growth--model parameter space.

\begin{figure}
     \begin{center}
     \includegraphics[width=\columnwidth]{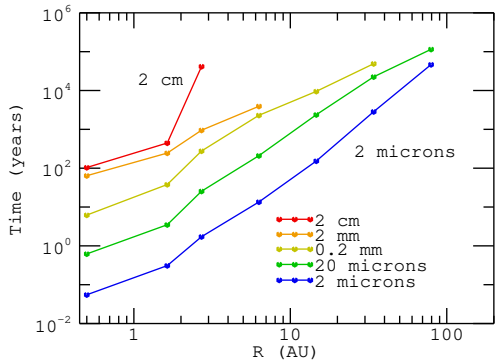}
     \caption[$\,\,$ Grain-growth timing within the fiducial disc]{Timing for the initiation of larger grains within our fiducial simulations for each of our radial zones.  These correspond to the more rapid of the two growth timescales produced by our cartoon grain-growth models, discussed in Appendix~\ref{AX_GrainGrowth}.  Points not plotted beyond a given radial position indicate that grains are not grown to that size within the disc lifetime.}
     \label{FTransport_Fgraingrowth}
     \end{center}
     \end{figure}

Figure~\ref{FTransport_Fgraingrowth}
plots the grain-growth timing employed within our fiducial disc simulations.  We run our two cartoon grain-growth models independently and initiate larger grain sizes according the most rapid growth timescale indicated.  Due to disc evolution and depletion, sometimes only one (or neither) model will produce grains of a given size within a given zone.  Our ten zones have characteristic radii spaced logarithmically from 0.5 to 1000~AU, and zone boundaries placed logarithmically between.  We simulate transport of 0.2~$\mu$m--2cm-sized grains, spaced per decade in grain size, running an ensemble of $2\times10^4$ grains for each grain-size transport simulation.  Grains initialized within a zone are placed randomly following the $t=t_0$ gas-mass distribution within that zone.

Each simulation particle tracks an assigned gas-equivalent mass, $m_\mathrm{GE}$, of solid material within the disc.  The gas-equivalent mass can be used to directly measure the local change in the dust-to-gas ratio (the dust enhancement factor, $E$) within the disc.  By adding up all gas-equivalent mass within a radial bin, $M_\mathrm{GE}=\sum m_\mathrm{GE}$, and comparing that to the disc-gas mass within that bin, $M_g$, we can calculate the locally-measured dust-to-gas enhancement factor within our simulations. 
\begin{equation}
   E
=
   \frac{M_\mathrm{GE}}{M_g} \,.
\label{ETransport_Efactor}
\end{equation}
$M_\mathrm{GE}=M_g$ for $E=1$ indicates that the local disc metallicity is unchanged from its initial (solar) value, whereas $E>1$ indicates an enhancement in the local dust-to-gas ratio.

Our simulations do not explicitly track refractory and icy material separately or include evaporation/condensation effects for grains crossing the snow-line.  Instead, the composition (and therefore relative mass of grains) is assumed to be that appropriate for the local disc-temperature conditions.  While the enhancement factor is, therefore, the fundamental output of our dust-transport simulations, it can be translated into a local value for the dust-grain surface density, $\Sigma_p$, separate from the disc gas.  To do this, we solve for $\Sigma_p$ at a given binned $\left(R,t\right) \rightarrow \left(i,j\right)$ using those assumed compositions and the more general definition of the enhancement factor,
\begin{equation}
   E_{i,j}
=
   \left(\frac{\Sigma_p}{\Sigma_g}\right)_{i,j} / \left(\frac{\Sigma_p}{\Sigma_g}\right)_\odot \,,
\label{ETransport_Efactorreal}
\end{equation}
where $\left(\Sigma_p/\Sigma_g\right)_\odot = Z_\odot$ refers to the initial (solar) dust-gas composition of the disc, the baseline metallicity.  $Z_\odot$ used in this work is taken from \cite{Lodders2003} and is broken into two components: 
\begin{equation}
   Z_\odot
=
   Z_0 Z_\mathrm{rel}\left(T\right) \,,
\label{LPC_Zsun}
\end{equation}
where $Z_0=0.0149$ is the total fraction of condensable material thought to be present in a solar-composition gas, and $Z_\mathrm{rel}\left(T\right)$ is the fraction of that material believed to actually be condensed based on the local gas temperature.  The values of $Z_\mathrm{rel}$ used are 
\begin{eqnarray}
   Z_\mathrm{rel}
=
   0.3289
&\mathrm{for}&
   T > 182 \mathrm{K} \,,
\nonumber \\
   Z_\mathrm{rel}
=
   0.7129
&\mathrm{for}&
   182 \mathrm{K} > T > 41 \mathrm{K} \,,
\nonumber \\
   Z_\mathrm{rel}
=
   1
&\mathrm{for}&
   T < 41 \mathrm{K} \,,
\nonumber
\end{eqnarray}
with transitions corresponding to the condensation of water ice (the snow line) and to the condensation of methane (and ammonia) ice.

Because initially all disc solids are contained within 0.2~$\mu$m grains, at $t=0$ all 0.2~$\mu$m grains are assigned $m_\mathrm{GE}=M_{D,0}/2\times10^4$.  When a new grain size, $2s_d$, is initialized within a radial zone, all $m_\mathrm{GE}$ values for grains 0.2~$\mu$m -- $2s_d$ presently within that zone are reassigned to follow an updated mass allocation for the grain-size distribution.  For this mass re-allocation, we assume a power-law distribution of grain sizes,
\begin{equation}
   \frac{dn_p}{ds_d}
\propto
   s_d^{q_s} \,,
\label{LPC_grainpowerlaw}
\end{equation}
where $n_p$ is the number density of grains of a given radius $s_d$, and $q_s$ is the power-law index of the distribution.   For the fiducial simulations, we take $q_s=-3.5$, corresponding to the size distribution of grains measured for the ISM \citep{MathisRumplNordsieck1977}.  However, observations linked to grain-growth within discs often predict flatter values of $q_s$ \citep[e.g.][]{Riccietal2011}, and in \S\ref{Results} we also report results that include $q_s=-2.5$ and -4, the later commonly used to depict a purely collisional size distribution, e.g., \cite{BaiStone2010b}.

To calculate the fraction of the total dust mass within a zone to be reallocated into a given grain-size bin, we follow \cite{Garaud2007}, where, for a general value of $q_s$,
\begin{equation}
   \frac{\rho_{p,s_d}}{\rho_p}
=
   \frac{\left(s_\mathrm{bmax}^{4+q_s} - s_\mathrm{bmin}^{4+q_s}\right)}
           {\left(s_\mathrm{max}^{4+q_s}-s_\mathrm{min}^{4+q_s}\right)} \,,
\label{LPC_massfraction}
\end{equation}
where $\rho_p$ is the local volume density of all solids, $\rho_{p,s_d}$ is the volume density of solids within size bin $s_d$, $s_\mathrm{min}$ and $s_\mathrm{max}$ are the lower and upper bounds, respectively, on the entire size distribution, and $s_\mathrm{bmin}$ and $s_\mathrm{bmax}$ are the bounds on a given size bin.  For $q_s=-4$,
\begin{equation}
   \frac{\rho_{p,s_d}}{\rho_p}
=
   \frac{\left[\ln \left(s_\mathrm{bmax}\right)-\ln \left(s_\mathrm{bmin}\right)\right]}
           {\left[\ln \left(s_\mathrm{max}\right)-\ln \left(s_\mathrm{min}\right)\right]}  \,.
\label{LPC_massfraction4}
\end{equation}
We define size bins logarithmically, so that, e.g., redistributing mass across 0.2~$\mu$m and 2~$\mu$m size bins ($s_d=0.1$ and 1~$\mu$m) has bin boundaries at $s_d=0.0316$, 0.316, and 3.16~$\mu$m.  Once we know the fraction of gas-equivalent mass to be portioned into each size bin, we divide it evenly among the particles of that size currently within the zone being updated.

Note that because our dust-transport trajectories are non-coagulative, tracking only a single grain size per model run, grain growth in our simulations in each radial zone occurs only when our grain-growth models indicate that the next largest grain size should be initiated.  At that time, all of the dust mass currently within that zone is summed and redistributed according to Equations~(\ref{LPC_massfraction}) or (\ref{LPC_massfraction4}), thereby updating the $m_\mathrm{GE}$ values of all particles (including those newly initiated) currently within that zone.  We therefore maintain a roughly equilibriated grain-size distribution within our disc during the early period of growth up to large-grain sizes.  However, as radial drift proceeds and preferentially depletes mass from the larger grain-size bins, this equilibration is not maintained.
Furthermore, we assume that larger grains will be built from grains only slightly smaller than themselves.  Therefore, in terms of mass reallocation, we do not initiate new, larger grains in a zone if radial drift has removed all grains of the next smallest size from that zone.

Without a mechanism to replenish the large-dust population, our simulations therefore produce a conservative estimate of the degree of dust-mass infall experienced within an evolving disk.  As (discussed in \S\ref{Results}) grain growth and the infall of large grains primarily leads to the draining of dust-mass from the disc, we expect that our simulations represent rough upper-limits on local dust-to-gas enhancements in the inner/main disc at times greater than the earliest disc-evolution timescales ($t \gtrsim 0.5$~Myr.)  For possible alternatives at earlier times, see \S\ref{Results_VaryModel}. 

Finally, we report the results for our dust-distribution simulations as a map of dust-to-gas enhancement, $E$, within radial and temporal space, $\left(R,t\right)$.  While our simulations report mass values for each radial grid-point (600 total) of our model disc and at each output time-step, usually $\delta t=200$~years, we present our results at lower $\left(R,t\right)$ resolution in order to reduce noise from our particle-ensemble representation of the dust distribution.  We present results using mass radially summed across three grid cells for 200 radial grid-points of results output, and averaged across usually 250 time-steps for a results $\Delta t = 5\times 10^4$~years.

     \section[Timescales and Disc-model Constraints]{Relevant Timescales and Disc-model Constraints}
     \label{Consequences}

\begin{figure}
     \begin{center}
     \includegraphics[width=\columnwidth]{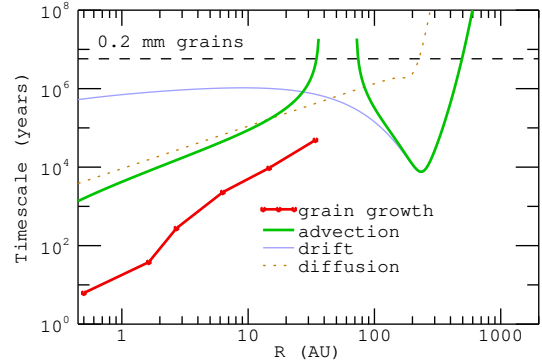}
     \caption[$\,\,$ Comparison of dust-transport and grain-growth timescales]{Comparison of dust-transport and grain-growth timescales for 0.2~mm grains within the fiducial disc model.  Timescales for inward advection, inward drift, and diffusion of grains are plotted for $t=10^5$~years of the fiducial disc model.  Timescales for grain-growth are from $t=0$ and include disc evolution.  There is a break in the plotted advection timescale where the dust flow is radially outward.  The dashed line references the disc lifetime.}
     \label{FConsequences_CompareTimescales}
     \end{center}
     \end{figure}

We begin by considering several timescales relevant to grain transport, which are plotted in
Figure~\ref{FConsequences_CompareTimescales}
for 0.2~mm grains at early times within the fiducial disc model.  The first is the time-scale to grow grains to a given size, for which we plot the initiation timing taken from the models shown in Figure~\ref{FTransport_Fgraingrowth}.  Next is the time-scale for grains to spiral inward onto the parent star.  This can be approximated as the distance to the star divided by the local inward-drift/drag velocity.  However, there are actually two inward velocities to consider.  The first is the head-wind drag velocity by itself, which is commonly used in such loss estimates applied to static disc models.  The second is the drag-advection velocity on the grain, calculated as $v_{srd}$ from Equations~(\ref{ETransport_svrda}) \& (\ref{ETransport_svrdb}) above, which includes the (mostly) inward gas flow resultant from disc evolution.  Within our model, the drift-only velocity, $v_{r,\mathrm{drift}}$, may be calculated with the same method as $v_{srd}$ only setting $v_{r,g}=0$.  The two inward-loss timescales are therefore
defined as
\begin{eqnarray}
   t_\mathrm{drift}
&=&
   - R / v_{r,\mathrm{drift}}
   \,,
\label{EConsequences_tdrift}
\\
   t_\mathrm{advect}
&=&
   - R / v_{srd}
   \,.
\label{EConsequences_tadvect}
\end{eqnarray}
The final timescale to consider is the timescale for turbulent/diffusive redistribution of the particle ensemble, which for Figure~\ref{FConsequences_CompareTimescales} we define as
\begin{equation}
   t_\mathrm{diffuse}
=
   R^2 / D_p
   \,.
\label{EConsequences_tdiffuse}
\end{equation}

From the timescales plotted in Figure~\ref{FConsequences_CompareTimescales}, one sees that there are several important trends to consider.
\begin{enumerate}
\item The shortest time-scale is that for the appearance of these 0.2~mm grains within the disc, out to at least a few tens of AU. At any one radius, then, it would be reasonable to assume that the initial phases of grain growth are effectively instantaneous. However, when we consider the disc as a whole, we find that even our lower limit on the growth time scale in the outer disc {\em exceeds} other time scales in the inner disc. This implies that by the time sub-mm grains have formed at tens of AU, we expect the major fraction of similar grains to have been lost from the inner few AU.
\item There is a large difference between the time-scales $t_\mathrm{drift}$ and $t_\mathrm{advect}$. The inward drift time scale, which would represent the flow of particles in an inviscid disc, is relatively long and almost constant with $R$. These features are consistent with the idea that radial drift results in a pile-up of solids in the inner disc. In an evolving disc, however, the advection time scale is substantially shorter everywhere inside 10~AU. Solids can therefore be lost via advection along with the gas, without having had time to pile-up.
\item Finally, the advection and diffusion timescales are similar, except in the outer disc where headwind-drag infall dominates. Turbulent redistribution of grains can therefore matter in the inner disc. Since turbulence represents a diffusive process, it will act to limit sharp radial gradients in particle concentration.
\end{enumerate}

For planetesimal formation, two basic properties of the background disc model are particularly important: the 
solid-gas coupling strength for particles of a given size, and the magnitude of the deviation from Keplerian 
rotation due to gas pressure gradients. These quantities enter into models for planetesimal formation via 
either streaming instabilities, or vertical precipitation followed by collapse. It is therefore useful to 
discuss how they vary in the gas disc model that we assume.

Local dust-gas coupling is parameterized using the local gas-drag stopping time on particles within the disc.  This depends on the size of the particle considered, but for characterizing the disc, we can plot the local particle-size corresponding to $\tau_s=\Omega_\mathrm{K} t_\mathrm{stop}=1$, which is a stopping time of particular interest to some theories of planetesimal formation.  In the Epstein-drag regime
\begin{equation}
   \tau_s
=
   \frac{s_d \rho_d \Omega_\mathrm{K}}{\rho_g v_\mathrm{therm}} \,,
\label{EConsequences_taus}
\end{equation}
where $s_d$ and $\rho_d$ are the dust-grain radius and internal density, $\rho_g$ is the local gas density, and $v_\mathrm{therm}$ is the thermal velocity of the gas particles.  At the midplane in a vertically-isothermal disc model, Equation~(\ref{EConsequences_taus}) translates to
\begin{equation}
   \tau_s
=
   \frac{\pi s_d \rho_d}{2 \Sigma_g}
   \,.
\label{EConsequences_tausIsothermal}
\end{equation}

\begin{figure}
     \begin{center}
     \includegraphics[width=\columnwidth]{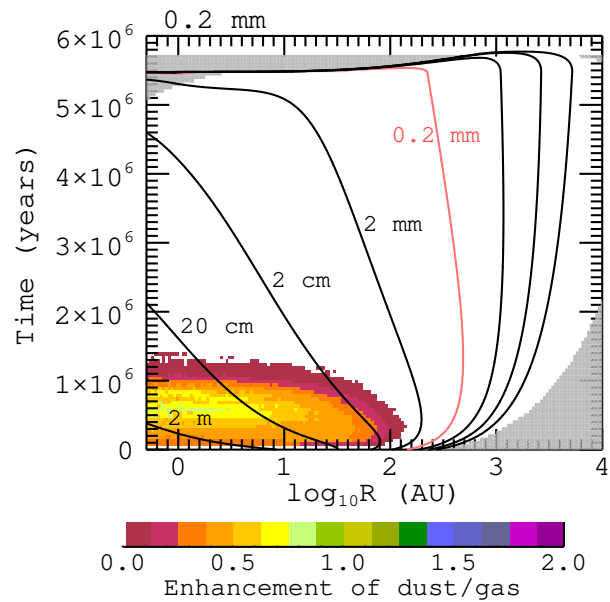}
     \caption[$\,\,$ $\tau_s=1$ contours overlying simulated 0.2~mm grain distribution]{$\tau_s=1$ grain-size contours ($2s_d=0.2 \mu$m--2 m) overlying our simulated distribution of 0.2~mm grains within the fiducial disc model.  Enhancement values of the distribution are scaled to the total simulated dust-grain distribution, of which 0.2~mm grains make up some fraction.  Note that $\tau_s=1$ contours follow $\Sigma_g$ contours in a vertically isothermal disc (Eq.~\ref{EConsequences_tausIsothermal}).}
     \label{FConsequences_tausV02mmDistr}
     \end{center}
     \end{figure}

In Figure~\ref{FConsequences_tausV02mmDistr},
we plot the $\tau_s=1$ contours for $\rho_d=1$~g~cm$^{-3}$ grains within the fiducial disc model, overlain on the 0.2~mm component of our simulated dust-grain distribution.  We have highlighted the contour corresponding to $\tau_s=1$ for 0.2~mm grains, and one can see that
the dust distribution within the disc is confined well interior ($\tau_s < 0.1$) to this contour.  This is consistent with the time-scales for radial drift presented in Figure~\ref{FConsequences_CompareTimescales} ($\tau_s=1$ occurs near 200~AU at the peak in the head-wind-drag effect), and also with the simulations of \cite{BrauerDullemondHenning2008}, though radial confinement in our fiducial simulations is about an order-of-magnitude greater than theirs, perhaps due in part to our approximate treatment of grain growth.  We find identical behavior (confinement to $\tau_s<0.1$ regions of the disc) for the simulated distributions of each dust-size component we have run, including only minor variation if the diffusivity ($1/Sc$) is varied by a factor of two.  Headwind drag therefore confines larger dust solids to the inner/main disc.

The precipitation mechanism of planetesimal formation explored by \cite{YoudinShu2002} is formally independent of the local dust size distribution (though grains do need to be large enough to settle toward the disc midplane, particularly if intrinsic disc turbulence is important for the vertical dust profile).  However, streaming-instability models for planetesimal formation find that particle clumping and collapse occurs specifically for particles that are marginally coupled to the gas motions \citep{Johansenetal2007,JohansenYoudin2007,BaiStone2010b}.  This means particles with
normalized stopping times, $\tau_s$, preferentially near 1 (though \cite{BaiStone2010b} report that grains as small as $\tau_s \sim 10^{-2}$ still participate in the streaming instability).  Because the headwind-drag effect dictates the rapid removal of grains from their $\tau_s \sim 1$ disc locations, it is hard to reconcile planetesimal formation models that require $\tau_s \sim 1$ with aerodynamic grain-transport theory.  Observationally, however, chondrules ($\sim$mm in size) do make up a large fraction of most meteorites \citep{CuzziDavisDobrovolskis2003}. Planetesimals in the inner region of the Solar Nebula may then have been largely built from mid-to-large sized grains (see, e.g., \cite{CuzziHoganBottke2010} for a discussion), and planetesimal formation models using such grain sizes may be well justified.

     \begin{figure}
     \begin{center}
     \includegraphics[width=\columnwidth]{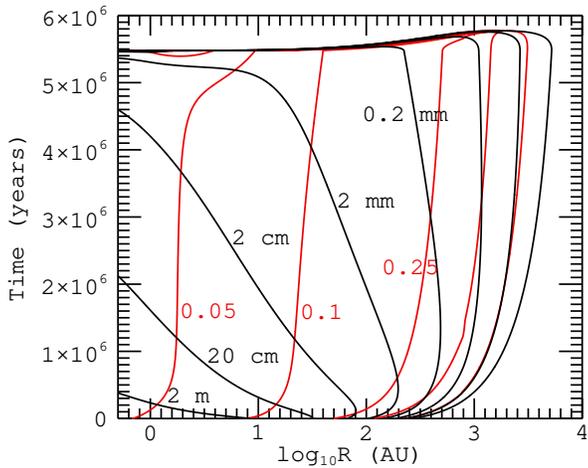}
     \caption[$\,\,$ $\Pi_\eta$ contours versus $\tau_s=1$ grain-size contours]{$\Pi_\eta$ contours {\bf (red)} overlying $\tau_s=1$ grain-size contours {\bf (black)} for the fiducial-disc model.  $\Pi_\eta$ contours are for values of 0.05, 0.1, 0.25, 0.5, and 1.  $\tau_s=1$ grain-size contours are spaced in decades 0.2 $\mu$m--2 m.}
     \label{FConsequences_PietaVtaus}
     \end{center}
     \end{figure}

     \begin{figure*}
     \begin{center}
     \includegraphics[scale=0.7]{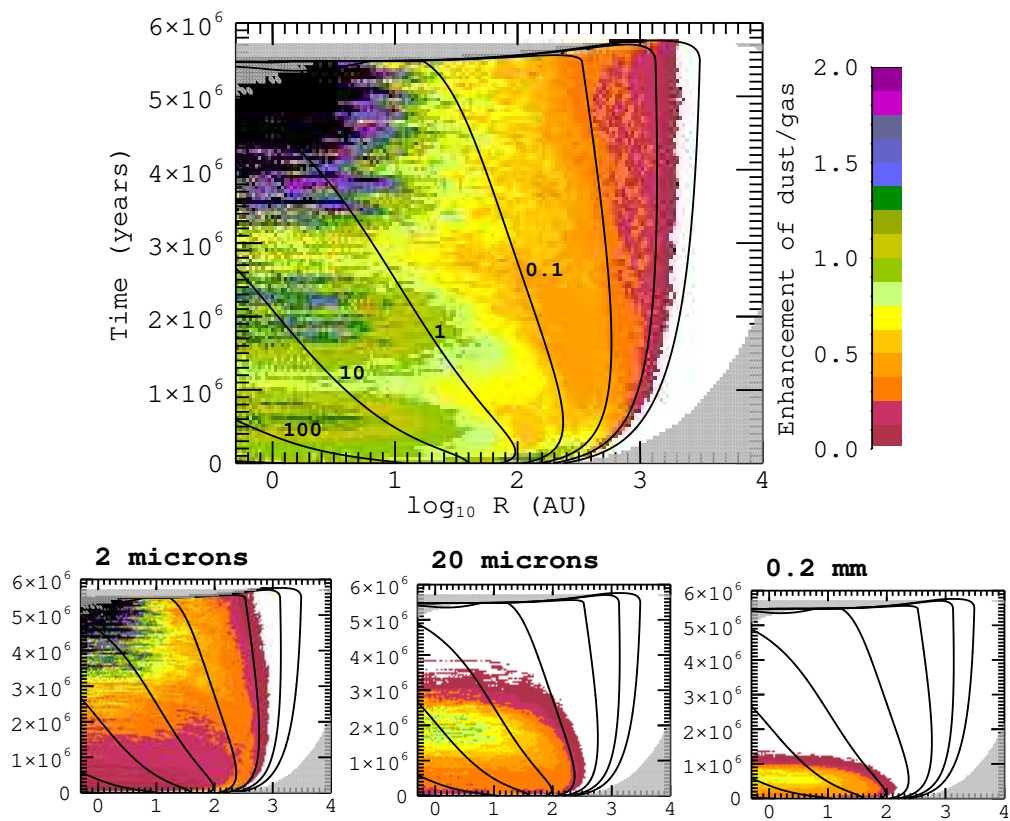}
     \caption[$\,\,$ Enhancement map and grain-size components]{Composite map of dust/gas enhancement factor (top panel) within the fiducial disc model and maps of some of the contributing grain-size bins (bottom panels).  Regions colored in black have a solids/gas enhancement factor of $\times 2$ or greater.  The black contours trace the evolving gas surface density, $\Sigma_g$, spaced by orders of magnitude between $10^{-4}$ and $10^3$ g cm$^{-2}$.  The grey regions mask out bins for which the particle statistics were too low to report dust/gas enhancement values.}  
     \label{FResults_FCompositeMap}
     \end{center}
     \end{figure*}

Finally, a requirement of planetesimal formation common to both precipitation and streaming-instability models 
(and consistent with the model presented in \cite{CuzziHoganBottke2010}) 
is that the local radial pressure gradient within the disc is small.  The radial pressure gradient can be parameterized via,
\begin{equation}
   \eta_{\delta \phi}
= 
   \frac{\left(v_\mathrm{K}-v_{\phi,g}\right)}{v_\mathrm{K}}
   \,,
\label{EConsequences_etadphi}
\end{equation}
since it is the source of the offset between gas and Keplerian orbital velocities.  The dependence of the \cite{YoudinShu2002} precipitation mechanism on low $\eta_{\delta\phi}$ is explicitly outlined in \S\ref{Results_VEnhancements}.  Some papers reporting simulations of streaming-instability clumping of particles instead use $\Pi_\eta \equiv \eta_{\delta\phi} v_\mathrm{K} / c_s$, or
\begin{equation}
   \Pi_\eta
=
   \frac{\left(v_\mathrm{K}-v_{\phi,g}\right)}{c_s}
   \,,
\label{EConsequences_pieta}
\end{equation}
and in Figure~\ref{FConsequences_PietaVtaus}
we plot $\Pi_\eta$ contours in red overlying the $\tau_s=1$ grain-size contours for the fiducial-disc model.  In \cite{BaiStone2010c}, the authors report simulations of particle clumping via the streaming-instability in which they vary $\Pi_\eta$ between 0.025, 0.05, and 0.1.  They find clumping requires $\tau_s \age 10^{-2}$  
for most particles and super-solar metallicity when $\Pi_\eta=0.05$, and a monotonic increase in required metallicity for increasing $\Pi_\eta$.  Comparing these results to the spatial distribution of $\Pi_\eta$ values within our disc model,  it is clear that even the purely gaseous properties of the disc work to disfavor planetesimal formation at large radii. It remains to be explored whether potentially significant dust-to-gas enhancements above solar can be produced closer in.

     \section[Results: The Global Distribution of Solids]{Results: The Global Distribution of Solids}
     \label{Results}

          \subsection[Fiducial Results]{Fiducial Results}
          \label{Results_Fiducial}

Figure~\ref{FResults_FCompositeMap} shows 
the simulated solids enhancement (change in the dust-to-gas ratio) as a function of $R$ and  $t$ within the model disc.   It is overlain with contours showing the evolution of the gas surface density.  The bottom panels of the figure plot contributions from the individual grain-size components that make up the total solid mass.  (Note that the green, scallop-shaped enhancement contours in the inner/main disc in the composite map are artifacts of the size-binning, matching the terminal-infall boundaries of successively larger grain sizes.)  In these maps, green areas represent no change from the $t=0$ dust-to-gas ratio, red areas represent depletion of dust relative to the gas, and black areas represent $\times 2$ enhancement of the dust or greater.

From the figure, we see that the fiducial-disc simulations do not produce large radial concentrations of dust in the main and inner discs until near the end of the disc lifetime.  In \S\ref{Results_VEnhancements}, we will compare these measured enhancement values to predicted requirements for the precipitation model of planetesimal formation.  Recall, however, that a common minimum requirement quoted to aid planetesimal formation is a doubling of the dust-to-gas ratio above solar metallicity, and that, as discussed in \S\ref{Introduction}, it is the first few million years of disc evolution that are of greatest interest in terms of bulk planetesimal formation. Potentially significant enhancements are not produced 
by our models 
during this phase of disc evolution.

Figure~\ref{FResults_FCompositeMap} illustrates several other characteristics of the simulations. 
At early times, a noticeable portion of the small-grain population is advected outward as the disc expands. However, inward drift of particles quickly sets in, and leads to the depletion of dust relative to the gas in the outer regions of the disc. In the inner disc, this influx of mass is in turn offset by concurrent loss (due to further inspiral) of most of the larger-grain population.   The net result is a significant loss of dust mass from the evolving disc.  

\begin{figure}
     \begin{center}
     \includegraphics[width=\columnwidth]{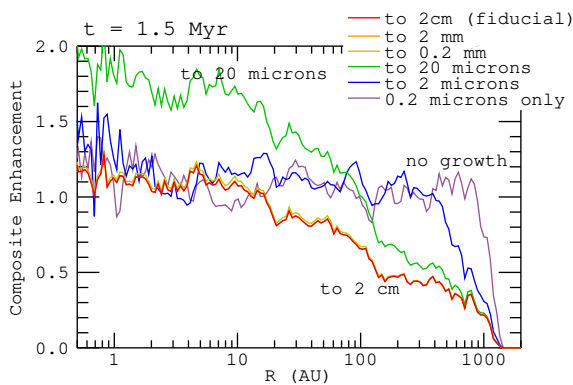}
     \caption[$\,\,$ $E$ at $t=1.5$ Myr for truncated grain-size distributions]{Measured enhancement of local dust/gas for the fiducial-disc model simulations at $t=1.5$ Myr varying the upper-limit on the grain-size distribution.  For no grain growth (0.2 $\mu$m grains only) the dust enhancement factor remains everywhere near unity.  Allowing larger grain sizes allows inward drift and some pile-up of the dust population.}
     \label{FResults_Egraingrowth}
     \end{center}
     \end{figure}

The significance of the mass contained within the larger grain sizes is depicted in Figure~\ref{FResults_Egraingrowth}
where, for $t=1.5$ Myr, we plot the enhancement factor produced when the results of our fiducial simulations are compiled (using sub-sets of the simulation output) for grain-size distributions truncated at successively smaller particle sizes.  Here, we see that grain-growth to roughly 20 $\mu$m sizes results in inward drift that depletes the outer disc of dust, and it is for grain-size truncation at $20 \mu$m that our simulations produce peak enhancement ratios within the inner/main disc.  Growth to larger sizes, however, subsequently depletes the inner disc of this dust-mass excess, bringing enhancement values back toward unity.

\begin{figure}
     \begin{center}
     \includegraphics[width=\columnwidth]{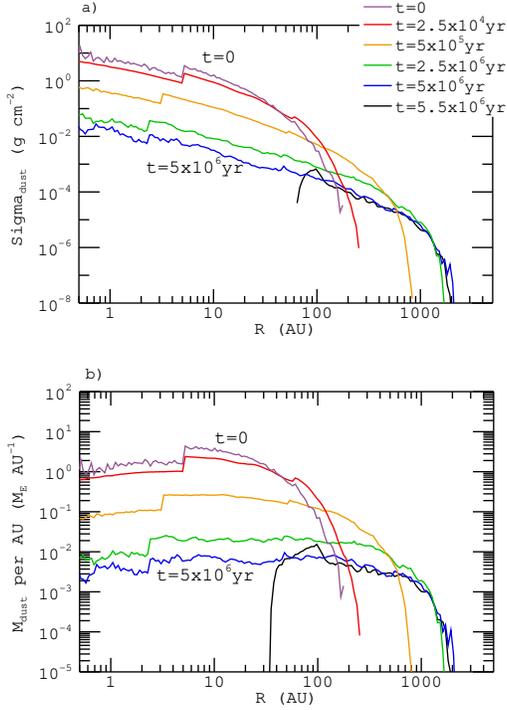}
     \caption[$\,\,$ Evolving dust-mass distributions for the fiducial simulations]{Time evolution of (a) the dust surface-density profile, and (b) the dust mass-per-radius distributions for the fiducial-disc simulations.}
     \label{FResults_FdustDistribution}
     \end{center}
     \end{figure}

Finally, we can solve for a representation of the local surface-density of dust particles, $\Sigma_p$, using Equations~(\ref{ETransport_Efactorreal}) and (\ref{LPC_Zsun}) in order to gain a sense of the diminution of solids available in an evolving, accreting disc as time progresses.  In Figure~\ref{FResults_FdustDistribution},
we plot the time evolution of $\Sigma_p$ and of the local dust-mass per AU within the fiducial-disc simulations.  The snow-line is clearly marked by jumps in the distributions at around a few AU.  At $t=0$, our fiducial disc has just over 100 Earth-masses of solid material spread throughout the entire disc, with the greatest concentration just beyond the snow line at around 5--10 AU.  As the disc evolves, the solids distribution spreads outward with the disc and decreases in total quantity, to 40 $M_\mathrm{Earth}$ at $t=5\times10^{5}$ years, and 10 $M_\mathrm{Earth}$ at $t=2.5$ Myr.

          \subsection[Varying Disc Model Parameters]{Varying Disc Model Parameters}
          \label{Results_VaryModel}

To quantify the role of radial dependence on dust/gas enhancement results through both the local maximum of grain sizes considered and the time delay in the formation of the largest grains, we present two simulations using alternative grain-growth model assumptions.  In the first, the initial particle-size distribution does {\em not} vary as a function of 
disc 
radius.  All particles of all sizes 
(0.2 $\mu$m -- 2 cm)  
are instead initiated at $t=0$ evenly distributed throughout the entire disc-gas-mass distribution.  This is similar to the radially independent size distribution used as an initial condition in the \cite{YoudinShu2002} calculations of dust pile-up via radial drift, except that we retain the simple power-law form of the size distribution with $q_s=-3.5$ throughout, while \cite{YoudinShu2002} considered a two-component distribution representing chondrules and matrix material.  

\begin{figure}
     \begin{center}
     \includegraphics[width=\columnwidth]{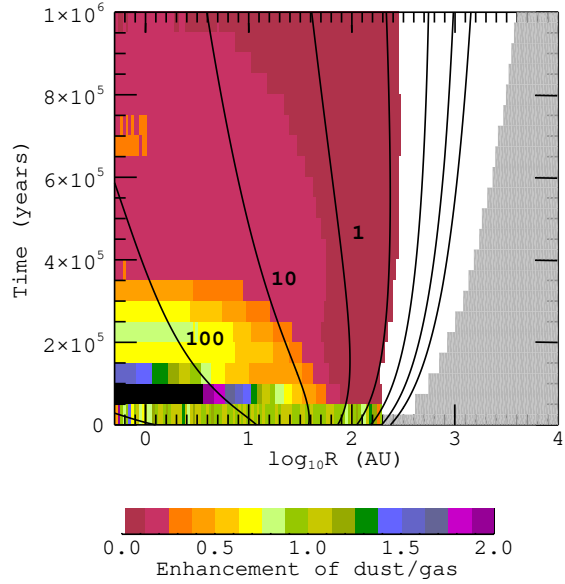}
     \caption[$\,\,$ Enhancement map for particles initiated at $t_0=0$ everywhere in the disc]{Enhancement map for a simulation using a radially independent initial grain-size distribution.}
     \label{FResults_tzeroMap}
     \end{center}
     \end{figure}

The results of our 
simulations using a radially independent grain-size distribution 
are plotted in Figure~\ref{FResults_tzeroMap}
and are distinct from those of our fiducial simulations.  Due to the large fraction of total dust mass now residing in large grain sizes, most of the dust-mass in the disc rapidly falls inward to small AU.  It is thus quickly lost onto the parent star, but not before creating a period of distinct enhancement in the dust-to-gas ratio in the inner disc.  The peak in this enhancement occurs at $t \sim 4\times 10^4$~years at about $\times 5$ solar (initial) metallicity.  (If we instead assume a maximum grain size of only 2~mm, the peak in the enhancement occurs at $t \sim 2\times10^5$~years.)  After this peak, disc metallicity declines swiftly, and by $t=5\times10^5$ years the metallicity across the whole disc is at a fraction its $t=0$ value.  While this extreme-case grain-size distribution does produce significant enhancements in the dust-to-gas ratio early in the disc lifetime, the time to reach this enhancement peak is shorter even than the expected time-scale for disc assembly ($\sim 10^5$ years). It is thus hard to envisage a disc formation scenario in which the pile-up seen in this instant-grain-growth-everywhere simulation could be physically realized.  Moreover, such a short window supplied by this simulation for planetesimal formation is in direct contradiction with meteoritical evidence implying large-body formation in the Solar System persisted for millions of years.  

\begin{figure}
     \begin{center}
     \includegraphics[width=\columnwidth]{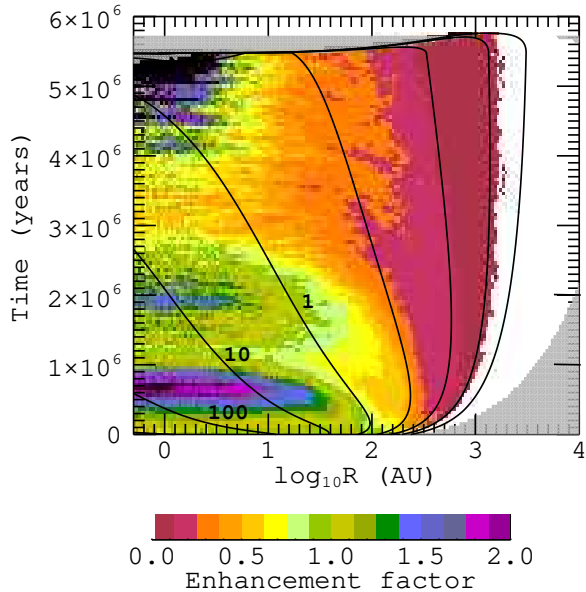}
     \caption[$\,\,$ Enhancement maps for tzero zoned]{Enhancement map for a simulation with all grains initiated at $t=0$, but with maximum grain sizes set by radial zone using our cartoon grain-growth model.}
     \label{FResults_tzero_zoned}
     \end{center}
     \end{figure}

In the second alternative simulation, we again assume that grain growth is instantaneous, initiating all particles at $t=0$.  However, we now return to the assumption that the largest grains will grow only in the inner, dense regions of the disc.  The results depicted in Figure~\ref{FResults_tzero_zoned} 
use the same simulation output as those of Figure~\ref{FResults_tzeroMap} but have been compiled selecting only those grains initiated in the zones appropriate for their size as defined by our cartoon grain-growth model, e.g., 2~mm grains initiated at $R\le 9.6$~AU. 
\footnote{Note: To compare to Figure~\ref{FTransport_Fgraingrowth}, 9.6~AU is the outer boundary of the radial zone with characteristic radius defined at 6.3~AU.}  
This simulation produces results similar to those of our fiducial simulations but with two notable exceptions.  First, the small-grain population is depleted relative to the fiducial simulations, producing a somewhat greater depletion of dust relative to the gas at large distances and at late times.  In the fiducial simulations, small grains have time to advect and diffuse into the massive outer disc before dust mass at smaller distances is converted into larger sized grains.

Second, there is a peak in the enhancement factor of just over $\times 2$ solar metallicity at small radii around $t=6\times 10^5$~years.  This peak corresponds to the terminal infall of the 0.2~mm sized grains, which are carrying a larger portion of the dust mass than in the fiducial simulations.  Assuming instantaneous grain growth also allows infall of these larger grains without delay from the further regions of the disc (out to $R=52$~AU).  While the 2~mm and 2~cm grain-size bins are also holding more dust mass than in the fiducial simulation and infall more swiftly than the 0.2~mm grains, their presence is too radially confined to produce an earlier peak in the dust/gas enhancement like that seen in Figure~\ref{FResults_tzeroMap}.  The enhancement peak produced in this simulation is timed more favorably to assist in planetesimal formation than the fully radially independent simulation, particularly under the potentially lower enhancement requirements of streaming-instability--type planetesimal formation.  However, its duration is still noticeably shorter than the meteoritical evidence would seem to indicate necessary.
 
The results of these simulations therefore seem to indicate that faster grain growth leads to a greater probability of dust pile-up in the inner regions of the disc, but that this effect is limited by the mass supplied in large grains initially a large distances from the star.  Furthermore, faster growth may still be insufficient to supply an extended period of dust/gas enhancement.  For the remainder of our simulations, we return to initiating particles radially and temporally based on our cartoon models of grain growth.

\begin{figure}
     \begin{center}
     \includegraphics[width=\columnwidth]{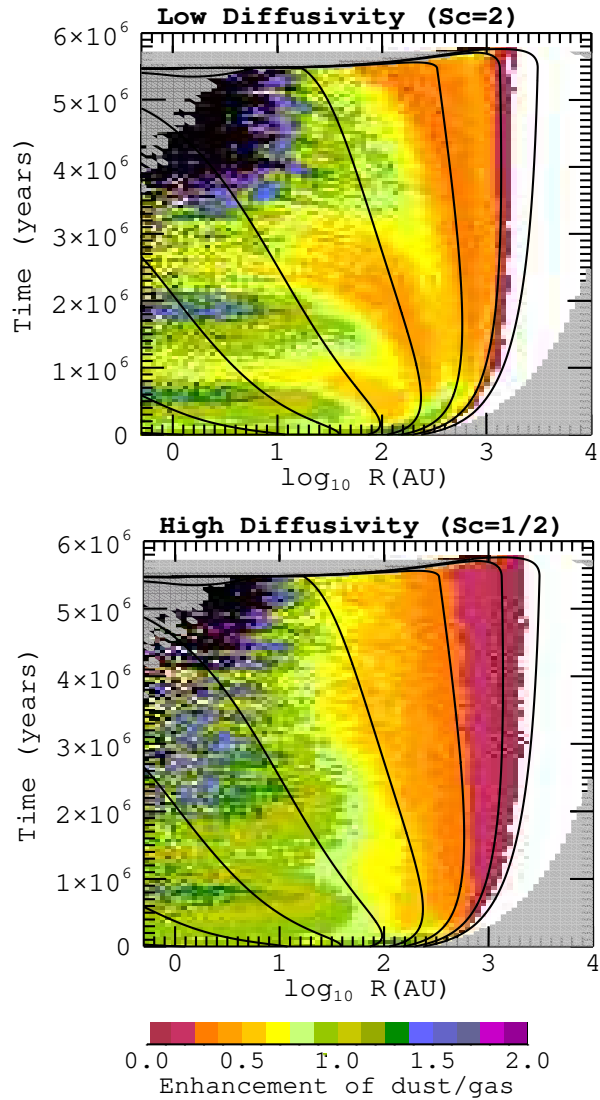}
     \caption[$\,\,$ Enhancement maps for simulations varying the diffusivity]{Enhancement maps from simulations run with $Sc=1/2$ and 2.  We use the fluid-dynamics convention that $Sc=\nu/D_g$, while retaining the local scaling for particle diffusivity based on grain size using Equation (\ref{ETransport_Dp}).  These simulations were run with output at lower resolution than in the fiducial setup and therefore compiled using six radial-grid-points per radial-bin and averaging 25 time-points for every $\Delta t=5\times 10^{4}$ years time-bin.}
     \label{FResults_diffusivity}
     \end{center}
     \end{figure}

Figure~\ref{FResults_diffusivity}
plots the enhancement distributions produced varying the Schmidt number, $Sc = \nu / D_g$ (inversely varying the gas-disc scaling for the particle diffusivity), by a factor of two in either direction.  We find the precise value of $Sc$ used to be unimportant to our results.  A lower diffusivity leads to greater segregation of the particle-size distribution and emphasizes the discrete particle sizes used for these simulations.  However, a real disc should have a mostly continuous particle-size distribution, and therefore the results for the global distribution of solids cannot be said to be qualitatively different from the fiducial model.  The higher-diffusivity case allows large grains to remain in the disc slightly longer than usual (all 20 $\mu$m grains are lost after 4.1 Myr rather than by 3.8 Myr as in the fiducial simulations), but again, this is insufficient to substantially alter the global distribution of solids.  

\begin{figure*}
     \begin{center}
     \includegraphics[width=\textwidth]{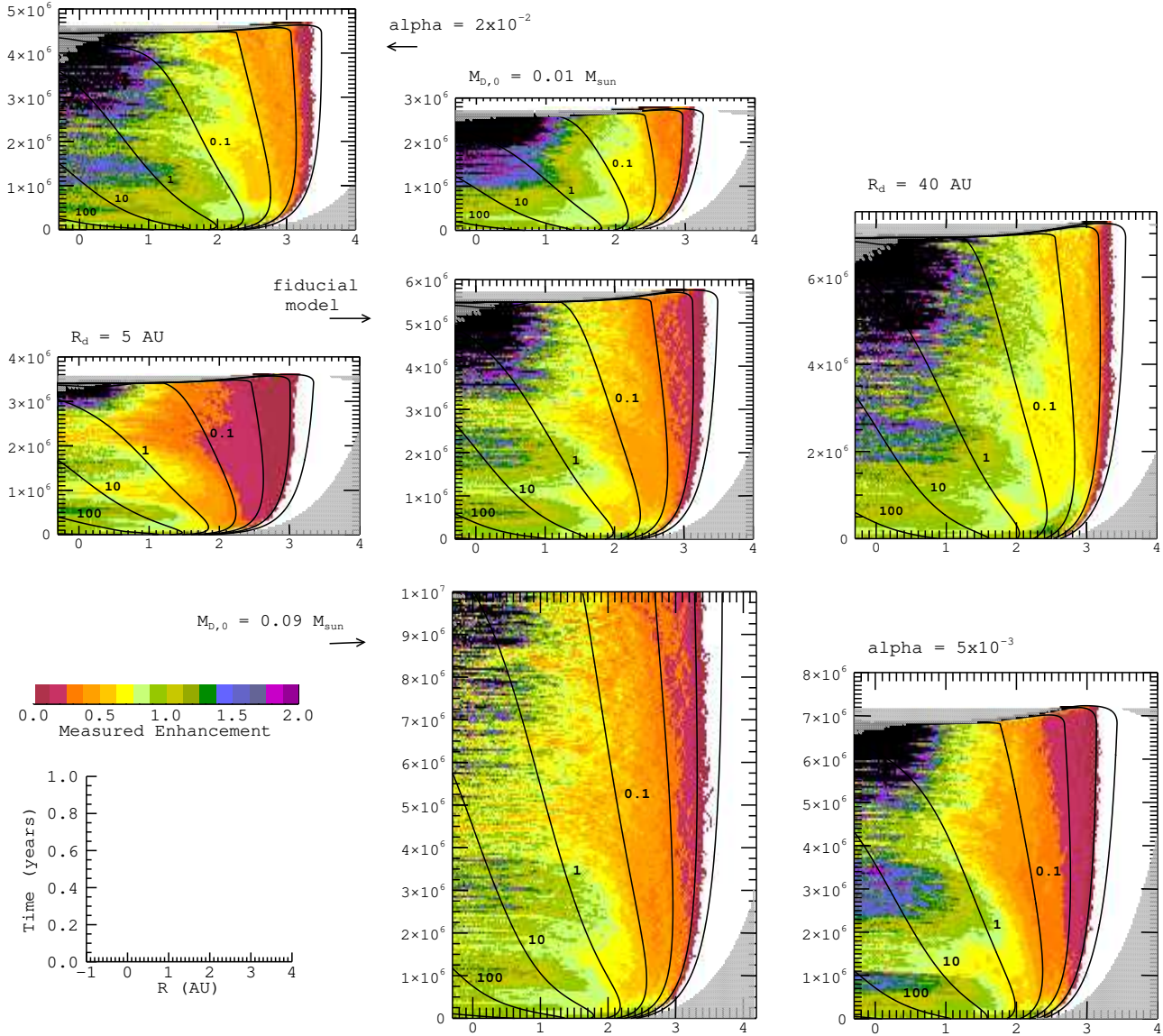}
     \caption[$\,\,$ Enhancement maps for simulations in different disc models]{Enhancement maps for simulations run in several different disc models, separately varying $R_d$ (horizontal), $M_{D,0}$ (vertical), and $\alpha$ (diagonal).  Fiducial-model map ($M_{D,0}=0.03 M_\odot$, $R_d=20$ AU, $\alpha=10^{-2}$) shown in the center panel.  Note that vertical scales are each set to the same years-per-inch to illustrate the varying rates of disk evolution.}
     \label{FResults_varyDiscs}
     \end{center}
     \end{figure*}

In Figure~\ref{FResults_varyDiscs},
we plot the composite enhancement maps for simulations run varying three disc-model parameters: the initial disc mass, $M_{D,0}$; the initial disc compactness, $R_d$, and the magnitude of the disc viscosity, $\alpha$.  Qualitatively, the patterns of evolving global enhancement are roughly constant across the different disc-model setups.  The infall of the smaller grain sizes causes depletion in the dust relative to the gas at the outer edge of the disc, while in the main disc enhancement values remain near unity for somewhere between 40 and 60\% of the total disc lifetime.  The initially more compact disc shows somewhat greater outer-disc depletion because a larger fraction of grains are processed to larger sizes at early times, while the disc using $\alpha=5\times 10^{-3}$ shows some of the enhanced segregation of our discrete particle sizes typical of the lower-diffusivity simulation in Figure~\ref{FResults_diffusivity} (as, in our transport model, lower $\alpha$ means lower disc viscosity {\em and} diffusivity).  However, the basic shape of dust-to-gas values within the disc as controlled by aerodynamic forces appears fairly universal.  The primary result, of these simulations then, is that a smoothly-defined, azimuthally-symmetric, evolving disc model tends to lose dust-mass smoothly inward onto the parent star at a rate that generally keeps pace with the loss of accreting disc-gas mass, and that variety in basic disc properties/parameters does not tend to produce special-case discs with respect to this process.

          \section[Comparison with Required Enhancement Factors]{Comparison with Required Enhancement Factors}
          \label{Results_VEnhancements}

In this section, we compare our simulated enhancement factors to those required by the \cite{YoudinShu2002} precipitation model of planetesimal formation. In the precipitation model, planetesimal formation is accomplished by the gravitational collapse of solids settled out to the disc midplane.  As outlined in \cite{Sekiya1998}, it assumes that the disc is quiescent, with no global turbulence to stir the particles upward, and that grains will settle out to the limit provided by the Kelvin-Helmholz shearing instability.  For midplane dust densities in this settled state above a certain mass threshold, the gravity of the dust sub-disc should dominate and lead to the collapse of solids at the midplane.  The criteria for this collapse, given in \cite{YoudinShu2002} Equation~(15), is
\begin{equation}
   \Sigma_{p,c}
=
   2 \sqrt{Ri_c} \eta_{\delta\phi} R \rho_g s\left(\psi\right) \,,
\label{LPC_YS02precip}
\end{equation}
where $\Sigma_{p,c}$ is the critical surface density of the solids for collapse, $Ri_c$ is the critical Richardson number defining the degree of dust settling prior to the onset of the Kelvin-Helmholtz instability, and $s\left(\psi\right)$ is a correction factor accounting for the self-gravity of the gas.  \cite{YoudinShu2002} define
\begin{equation}
   \eta_{\delta\phi}
\equiv
   - \frac{\left(\partial P/\partial R\right)}{2 \rho_g R \Omega^2} \,,
\label{LPC_etaphiYS02}
\end{equation}
and 
\begin{eqnarray}
   s\left(\psi\right)
&\equiv&
   \left(1+\psi\right)
   \ln \left[ \frac{\left(1 + \psi + \sqrt{1+2\psi}\right)}{\psi} \right]
-
   \sqrt{1+2\psi} \,;
\nonumber \\
   \psi
&\equiv&
   \frac{4\pi G \rho_g}{\Omega_\mathrm{K}^2} \,.
\label{LPC_spsiYS02}
\end{eqnarray}
For a gas density of $\rho_g \approx \Sigma_g / 2 H_g$,
where $H_g = c_s\,R / v_\mathrm{K}$ in a vertically isothermal disc model: 
\begin{equation}
   E_\mathrm{precip}
\ge
   \frac{ \left(\Sigma_p/\Sigma_g\right)_c }{ \left(\Sigma_p/\Sigma_g\right)_\odot}
=
   \sqrt{Ri_c} \frac{v_\mathrm{K}}{c_s} \frac{\eta_{\delta\phi} s\left(\psi\right)}{Z_0 Z_\mathrm{rel}\left(T\right)} \,,
\label{LPC_Eprecip}
\end{equation}
using the conventions for $Z_0$ and $Z_\mathrm{rel}\left(T\right)$ defined in \S\ref{ParticleTransport_Distribution}.  One expects that any degree of turbulence beyond the Kelvin-Helmholz effect should stir the particle distribution to greater vertical heights than those assumed in this $Ri_c$ scenario; certainly, the 
$\alpha = 10^{-2}$ 
used in our fiducial disc-evolution models suggests that a significant fraction of the small dust grains should be vertically well-mixed with the gas.  However, assuming that grains could settle out to their Kelvin-Helmholtz limit allows us to calculate the absolute minimum enhancement-factor required for precipitation of planetesimals to proceed
in that assumed non-turbulent environment.  

The precipitation-collapse criterion is separate from the simpler gravitational-collapse criterion using the Toomre-$Q$ parameter, $Q_p < 1$.  From \cite{YoudinShu2002},
\begin{eqnarray}
   Q_p
&\approx&
   \frac{\Omega^2 H_{p,c}}{\pi G \Sigma_{p,c}} \,;
\label{LPC_YS02Qp} \\
   H_{p,c}
&\equiv&
   \sqrt{Ri_c} \eta_{\delta\phi} R h\left(\psi\right) \,;
\nonumber \\
   h\left(\psi\right)
&=&
   \sqrt{1+2\psi} - \psi \ln\left[\frac{\left(1+\psi + \sqrt{1+2\psi}\right)}{\psi}\right] \,.
\nonumber
\end{eqnarray}
The minimum enhancement criterion for $Q$-collapse is then
\begin{equation}
   E_{Q_p}
\ge
   \sqrt{Ri_c} \frac{ v_\mathrm{K}^2 \eta_{\delta\phi} h\left(\psi\right)}
                              {\pi G R \Sigma_g Z_0 Z_\mathrm{rel}\left(T\right)} \,,
\label{LPC_EQp}
\end{equation}
which tends to dominate in the outer regions of massive, cold discs.

While the critical Richardson number is commonly set to $Ri_c=1/4$, recent numerical simulations by \cite{Leeetal2010} find the onset of Kelvin-Helmholtz instability occurring at
\begin{equation}
   Ri_c
\approx
   \frac{\epsilon_{\rho,0}}{36} \,,
\label{LPC_LeeRic}
\end{equation}
where $\epsilon_{\rho,0}$ is the current dust-to-gas volume-density ratio at the disc midplane.  $\epsilon_{\rho,0}$ can be written as $\epsilon_{\rho,0} \approx \left(\Sigma_p/\Sigma_g\right) / \left(H_p/H_g\right)$ and, if we continue to pursue to limit of maximum settling so that  $H_p$ follows from $H_{p,c}$ in Equation-set~(\ref{LPC_YS02Qp}), we can write $Ri_c$ from Equation~(\ref{LPC_LeeRic}) in terms of the current dust-to-gas enhancement and a $Ri$ value defining the current degree of settling.  Absolute maximum settling should therefore occur to $Ri_c$(Eq\ref{LPC_LeeRic})$=Ri$(Eq\ref{LPC_YS02Qp})$=Ri_c'$, or
\begin{equation}
   Ri_c'
=
   \left(\frac{c_s E Z_0 Z_\mathrm{rel}\left(T\right)}
                  {36 v_\mathrm{K} \eta_{\delta\phi} h\left(\psi\right)}
   \right)^{2/3} \,.
\label{LPC_LeeRicofE}
\end{equation}
Below, we compare our simulated dust enhancements to precipitation/collapse requirements using $Ri_c=Ri_c'$, so that Equations~(\ref{LPC_Eprecip}) \& (\ref{LPC_EQp}) become
\begin{equation}
   E_\mathrm{precip}
\ge
   \frac{v_\mathrm{K} \eta_{\delta\phi} s\left(\psi\right)^{3/2} }
          {6 c_s Z_0 Z_\mathrm{rel}\left(T\right) \sqrt{h\left(\psi\right)} } \,,
\label{LPC_Eprecip2}
\end{equation}
and
\begin{equation}
   E_{Q_p}
\ge
   \frac{\sqrt{c_s}\,v_\mathrm{K} \eta_{\delta\phi} h\left(\psi\right)}{6 Z_0 Z_\mathrm{rel}\left(T\right)}
   \left(\frac{\Omega_\mathrm{K}}{\pi G \Sigma_g}\right)^{3/2} \,.
\label{LPC_EQp2}
\end{equation}
This choice results in a slight decrease (up to a factor of two less) in minimum dust enhancements required in the inner/main disc to produce planetesimal formation relative to the \cite{YoudinShu2002} calculations, which use $Ri_c=1/4$.

\begin{figure}
     \begin{center}
     \includegraphics[width=\columnwidth]{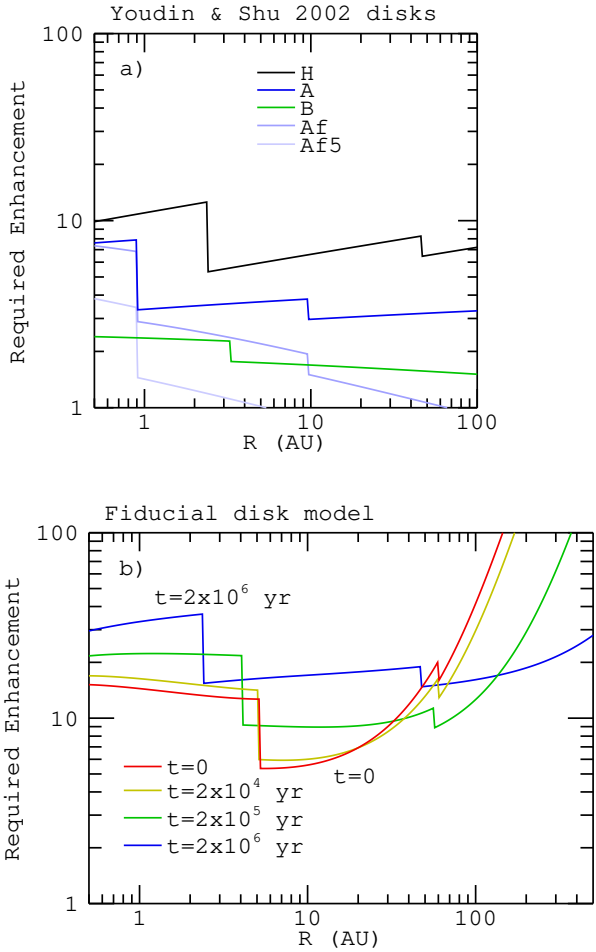}
     \caption[$\,\,$ Required Enhancements assuming $Ri_c = Ri_c'$ from \protect\cite{Leeetal2010}]{Required Enhancements for midplane precipitation using the results of \protect\cite{Leeetal2010} and Equation~(\ref{LPC_LeeRicofE}) to calculate $Ri_c$ for the discs of \protect\cite{YoudinShu2002} and our fiducial-disc model.  In our fiducial disc at $t=0$, the snow-line minimum value is $E_\mathrm{precip}=5.35$.}
     \label{FResults_ERicprime}
     \end{center}
     \end{figure}

Figure~\ref{FResults_ERicprime}
plots the required enhancement factors for the disc models discussed in \cite{YoudinShu2002} (panel-a) and for our fiducial disc at several times (panel-b).
In Table~\ref{LPC_YS02Vminetable}, 
\begin{table}
   \begin{center}
   \begin{tabular}{|c|c|c|c|c|}
   \hline
      {\bf model}
   & {\bf $\Sigma_\mathrm{AU}$ (g cm$^{-2}$)}
   & {\bf $-$q$_\Sigma$}
   & {\bf T$_\mathrm{AU}$ (K)}
   & {\bf $-$q$_T$}
   \\ \hline
      
   & 
   & 
   & 
   & 
   \\
      H
   & 1700
   & 3/2
   & 280
   & 1/2
   \\
      A
   & 1700
   & 3/2
   & 170
   & 0.63
   \\
      B
   & 1700
   & 3/2
   & 100
   & 3/4
   \\
      Af
   & 1700
   & 1
   & 170
   & 0.63
   \\
      Af5
   & 8500
   & 1
   & 170
   & 0.63
   \\
      
   & 
   & 
   & 
   & 
   \\
      {\bf fiducial}
   & {\bf 1180}
   & {\bf 1}
   & {\bf 280}
   & {\bf 0.6$\rightarrow$0.5}
   \\ \hline
   \end{tabular}
   \end{center}
\caption[Disc parameters for \protect\cite{YoudinShu2002} and our fiducial disc model]{Parameters for disc models used in \protect\cite{YoudinShu2002} versus rough equivalents for our fiducial disc near $t=0$.  In general, $\Sigma_g\left(R\right) = \Sigma_\mathrm{AU} \left(R/1\mathrm{AU}\right)^{q_\Sigma}$ and $T\left(R\right) = T_\mathrm{AU} \left(R/1\mathrm{AU}\right)^{q_T}$.  "H" designates the minimum-Solar-Nebula model of Hayashi-1981 (see \protect\cite{ArmitageBook} pp 4--5).}
\label{LPC_YS02Vminetable}
\end{table}
we list the main parameters defining the different \cite{YoudinShu2002} disc models, as well as the values roughly corresponding to the inner regions of our fiducial disc at early times.  (\cite{YoudinShu2002} use simple power-law disc models while our models have an exponential tail-off of surface-density in the outer disc.)

Figure~\ref{FResults_ERicprime} demonstrates several interesting features of these enhancement requirements.
\begin{itemize}
\item First, note that our fiducial model calls for remarkably high $E_\mathrm{precip}$ in the outer disc, particularly at early times.  This is due to the strong dependence of $E_\mathrm{precip}$ on $\eta_{\delta\phi}$.  When the radial pressure gradient is steep, as  near the outwardly expanding disc edge, shearing between the dust and gas is strong, requiring large dust-to-gas ratios to be overcome.  This limiting requirement is shared by some other proposed models of planetesimal formation, such as the streaming-instability model, as discussed in \S\ref{Consequences}.
\item  Second, while the \cite{YoudinShu2002} paper de-emphasizes the roles of the disc gravity, represented by $s\left(\psi\right)$, the time-series for $E_\mathrm{precip}$ of our fiducial model, particularly in the inner disc, demonstrates the importance of the local disc mass for meeting dust/gas-precipitation requirements.  $s\left(\psi\right)$ (and therefore $E_\mathrm{precip}$) becomes large, when $\psi$ is small, which may occur for either a low 
local-disc 
surface density, or for a locally hot disc.
\item Finally, the $E_\mathrm{precip}$ curves for both sets of disc models emphasize the importance of the ice lines, particularly the snow line, as places of potential local minimum in metallicity requirements for planetesimal formation.  While our particle-transport model does not treat evaporation and vapor-transport effects, the simulations of \cite{CieslaCuzzi2006} do show an additional peak in the dust population just outside the snow line due, at least partially, to local concentration of diffused and recondensed vapor.  This also corresponds to a peak in the disc opacity and heating, however, so it is not clear what the degree of advantage the region just outside the snowline lends to planetesimal formation via gravitational precipitation.
\end{itemize}

\begin{figure*}
     \begin{center}
     \includegraphics[scale=0.8]{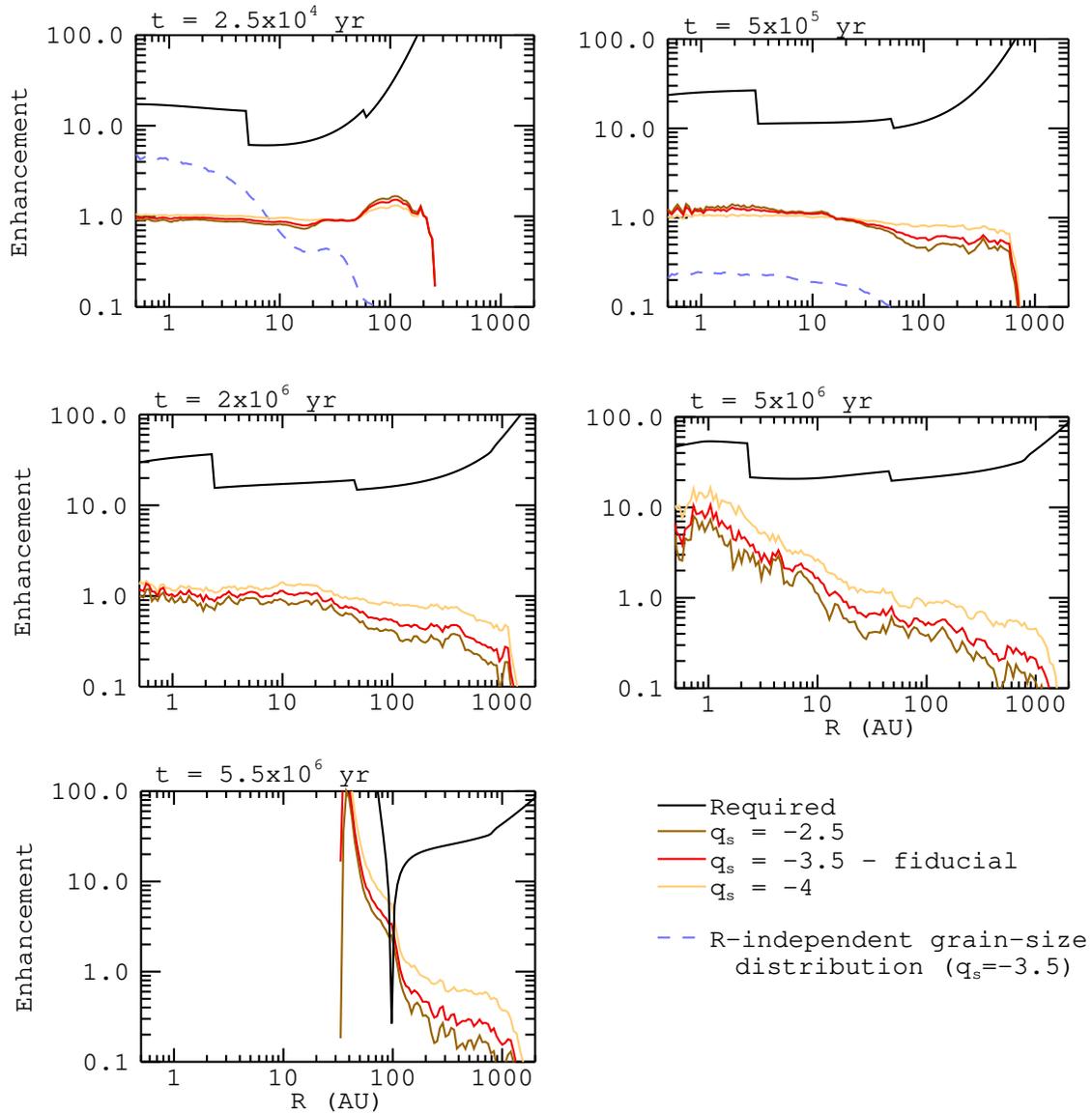}
     \caption[$\,\,$ $E_\mathrm{precip}$ in the fiducial model versus measured $E$ for several $q_s$ values]{Measured enhancement values for simulations in the fiducial-disc model at several times and using several particle-size distributions.  Compared to the enhancement factor required for precipitation/collapse as given by Equations~(\ref{LPC_Eprecip2}) \& (\ref{LPC_EQp2}).  The dashed curves plot the simulated distribution using the $t_0=0$, $R$-independent grain-size distribution discussed in \S\ref{Results_VaryModel}, which in the left-top-most panel is plotted at the peak in enhancement, $t=4\times 10^4$ years.}
     \label{FResults_Enhqs}
     \end{center}
     \end{figure*}

In Figure~\ref{FResults_Enhqs},
we compare minimum enhancement factors required for planetesimal precipitation to enhancement results from our simulations compiled using three different assumed particle-size distributions. From this figure, it is clear that the size distribution used to compile our simulation results has little impact on the conclusions of the fiducial runs.  Radial drift of particles does not produce enhancement sufficient to lead to planetesimal formation via precipitation for this model disc.  The exception occurs at the very end of the disc lifetime when photoevaporative clearing of the disc leads to an outward-sweeping pressure-maximum point at large AU where $\eta_{\delta\phi}$ goes to zero and $E_\mathrm{precip}$ drops to small values.

The first two panels of Figure~\ref{FResults_Enhqs} also include the results of the 
simulation 
discussed in \S\ref{Results_VaryModel} where all particle sizes were initiated at $t=0$ and distributed throughout the entire disc gas mass, with no radially-dependent grain-growth restrictions.  While this simulation yields a distinct peak in the dust-to-gas enhancement early on in the inner disc, it is still less than the 
minimum 
enhancement required by the precipitation model.

\begin{figure}
     \begin{center}
     \includegraphics[width=\columnwidth]{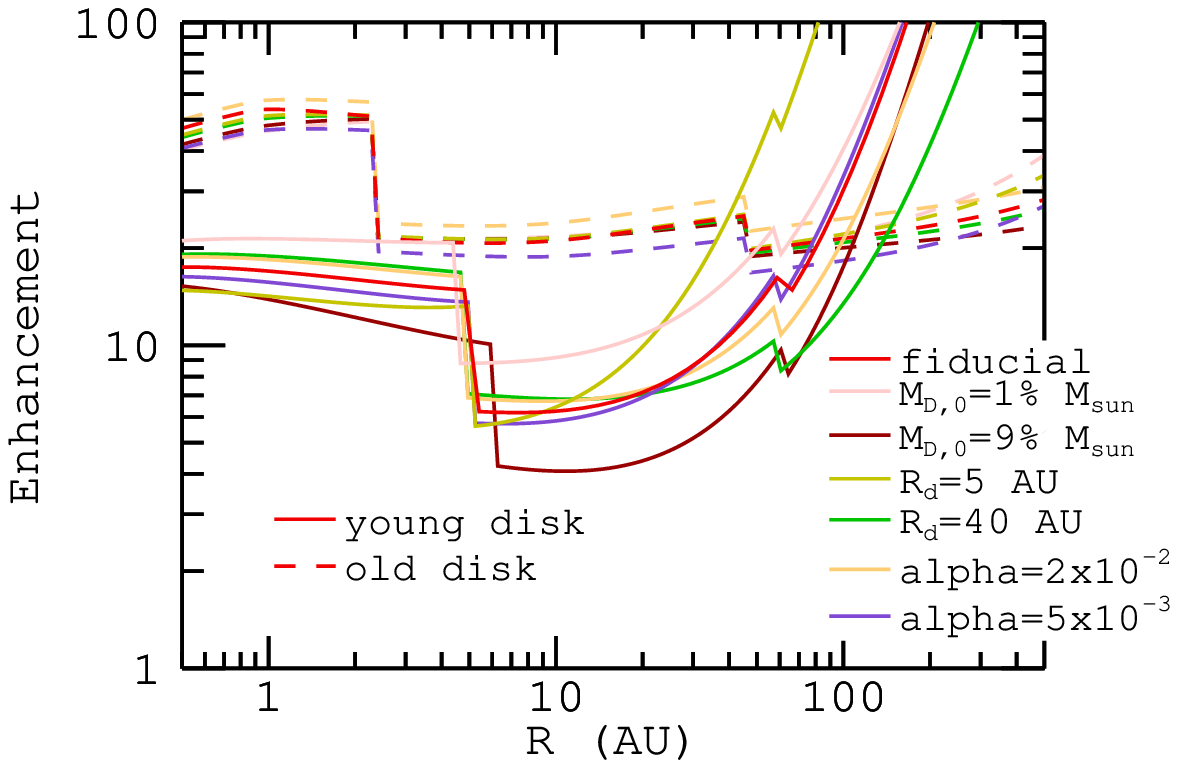}
     \caption[$\,\,$ $E_\mathrm{precip}$ in other model discs]{$E_\mathrm{precip}$ values near the beginning ($t=2.5\times 10^4$ years) and end (before the onset of photoevaporation) of the disc lifetime for the various evolving-disc models employed.  In the fiducial disc $M_{D,0}=0.03 M_\mathrm{sun}$, $R_d=20$~AU, and $\alpha=10^{-2}$.  Note that the onset of photoevaporation and disc clearing is controlled by the disc accretion rate.}
     \label{FResults_EreqDiscs}
     \end{center}
     \end{figure}

Finally, while the simulated enhancement distributions vary little between different evolving-disc models (as shown in Figure~\ref{FResults_varyDiscs}) the enhancement factors required by the precipitation model should vary according to factors such as disc mass.  In Figure~\ref{FResults_EreqDiscs},
we plot the enhancement criteria for precipitation near the beginning and end of the disc lifetime for the various discs used in generating Figure~\ref{FResults_varyDiscs}.  While more massive discs, in particular, produce lower minimum enhancement requirements, these requirements remain generally high relative to the dust-to-gas enhancements produced in our simulations.

     \section[Comparison to Disc Observations]{Comparison to Disc Observations}
     \label{CompareObservations}

One of the most general results from our simulated dust distributions is that we predict the outer-most edges of a disc should be depleted in dust relative to the gas.  Furthermore, we predict that the dust that does reside in the outer reaches of the disc should be fine-grained in order to remain mixed with such tenuous gas.  It is not yet clear from observations whether the outermost regions of a disc tend to be gas rich, though our prediction is consistent with some observations that find smaller disc radii when examining the dust versus the gas signatures of a disc \citep{HughesAMetal2008}.  As good observations at the outer edges of a disc require high sensitivity, a full test of outer-edge disc metallicities awaits more sophisticated facilities, such as ALMA.

However, there are observations of dust grain-size distributions within discs around some stars that are inconsistent with the dust-distribution simulations we have presented.  
Our simulations predict 
a global loss of mm-sized and larger grains from the disc within half a million years (even in our simulation of \S\ref{Results_VaryModel} where grains of all sizes were placed out to the largest distances within the disc).  However, observations in the (sub-)mm bands see evidence for mm-sized grains, not just within discs, but out to 100 AU or farther within those discs \citep{Testietal2003,Riccietal2010b}.  The cartoon grain-growth model we use to constrain the radial extent of simulation particles does not allow mm-sized grains to be initiated beyond 22 AU an any of the disc models we've run, and the inward flow from both accretion and headwind drag certainly precludes any simulated mm grains from reaching 100 AU.  Furthermore, even in the simulations of \cite{HughesArmitage2010} that considered grain transport (using the same fiducial disc model) within an {\em outward} flow of disc gas at the midplane, less than 6\% of mm-sized grains initiated between 0.5 and 10~AU reached even 25 AU, and then only for a brief period of time before headwind drag forced them inward once more.

That headwind drag tends to bar large grains from the outer disc, in contradiction with observations, is well-known 
\citep[e.g.][]{Weidenschilling1977a,TakeuchiLin2005,Braueretal2007}. \cite{Braueretal2007}
specifically focus on attempting to solve the radial-drift problem by including 
increasingly sophisticated 
physics into their transport calculations, including dust settling and collective effects, and angular momentum exchange between a dust sub-layer and the main gas disc.  The authors find these effects in general insufficient to slow radial drift of mm-sized grains to velocities compatible with their observed persistence in discs that are generally 1 Myr old or older.

The apparent inability of dust transport models to match observations poses a puzzle, since the basic one-dimensional description of the {\em gas} disc evolution appears to be a reasonably good fit to statistical measures of the  population such as disc sizes, lifetimes, and accretion rates.  It is possible that the axisymmetric gas disc model, while globally descriptive, is qualitatively inadequate as a background model for studies of particle transport.  One way in which this could happen is if local or non-axisymmetric structures are integral to accurately describing the large-scale distribution and transport of large dust grains.  
Otherwise, explaining outer-disc, large-grain populations with current aerodynamic disc theory seems to require such extreme scenarios as a collisional population of outer-disc planetesimals (of unknown origin), or else {\em very} massive outer gas discs of {\em very} low metallicity (due to the sustained depletion of large grains observed to be in residence) and currently unknown impact on global disc evolution.  

     \section[Discussion and Conclusions]{Discussion and Conclusions}
     \label{DiscussionConclusions}

In this paper, we have presented simulations of the global (re)distribution of dust within protoplanetary discs. Our models include evolution of the gas-disc surface-density profile, aerodynamic advection and diffusion of the dust-particle ensemble within the gas, and simulated grain-growth constraints confining the appearance of the largest grain sizes to the inner/main regions of the disc.  We find that the global distribution of solids within a 1D evolving disc model follows a fairly common pattern, and does not depend strongly on the values of parameters characterizing the initial disc, or on those describing the turbulence within it. Specifically,

\begin{enumerate}
\item Large, mm or cm-sized particles are lost rapidly onto the the parent star (within 0.5 Myr).
\item Growth of dust grains up to tens-of-microns size and larger leads to the depletion of dust relative to the gas toward the outer regions of the disc (beyond $\sim$100 AU).
\item The dust-to-gas ratio within the main and inner discs remains near solar (its initial value) for at least the first 40--60\% of the disc lifetime.
\end{enumerate}
In general, grains of a given size are confined to regions of the disc interior to at least $\tau_s \leq 10^{-1}$ and experience accelerated infall as $\tau_s$ ($\propto \Sigma_g$) contours near the outer boundary of that dust-population distribution evolve inward.

In \cite{YoudinShu2002}, the authors present calculations of dust-to-gas enhancement factors resulting from radial drift of dust grains within a static disc, and find a steady increase in dust-to-gas ratios at small disc radii with time.  This is in marked contrast to our own simulation results, which do not support the radial-drift hypothesis as a means to increase the metallicity in the inner disc in time for bulk planetesimal formation.  The difference between the two calculations is primarily the difference between a purely static- and an evolving-disc scenario.  As outlined by the drift/advection timescales presented in Figure~\ref{FConsequences_CompareTimescales}, a static disc produces ever-slower inward drift as grains approach the central star, providing a natural environment for grain pileup.  An evolving disc, however, will advect material steadily inward onto the parent star.  Furthermore, as the disc evolves and thins, drift due to headwind drag is accelerated, thus limiting the slowing of inward drift possible, even presuming a zone of the disc were to be free of an accretion/advection flow.

While our simulations do not paint a picture of large-scale increases to disc metallicity within an evolving protoplanetary disc, they do suggest a state of long-term continuity of conditions, perhaps for the first several million years of disc lifetime.  This is in keeping with observations that suggest the processing of solids in discs starts early \citep{Kwonetal2009,Riccietal2010a} but extends late, as evidenced by the spread in ages measured for chondrules embedded within meteorites \citep{Amelinetal2002}.  Furthermore, that our models predict similar conditions across a range in disc-model parameter space is in keeping with the apparent ubiquity of large-body processing within discs around other stars.  The fraction of stars observed to have circumstellar debris discs (with scattered-light levels around a thousand times brighter than that of our own zodiacal dust and Kuiper belt) peaks at around 50\% for 20 Myr old B and A stars \citep{CurriePlavchanKenyon2008}. 

We have failed to find circumstances in which global redistribution of solids, due to radial drift, results in 
substantial enhancements in the dust-to-gas ratio anywhere within an evolving disc. If large enhancements are in fact  a prerequisite for planetesimal formation, it may instead be necessary to appeal to local or non-axisymmetric  structures to provide them, 
although the effect of such structures may also be constrained by the headwind-drag produced lack of $\tau_s\sim1$ particles, discussed in \S\ref{Consequences}.  
Local disc structures that could prove important for dust concentration and/or transport include disc-opacity transitions, such as at the snow-line, spiral arms produced within gravitationally unstable discs \citep{Rice2006}, and vortices within the flow of disc gas. In  discs subject to the magnetorotational instability, some simulations suggest that MRI turbulence produces local structure sufficiently long lived to concentrate dust particles \citep{JohansenYoudinKlahr2009}. Determining which, if any, of these processes is important for particle transport and planetesimal formation will require an understanding not just of the strength of disc turbulence, but also of its detailed physical nature.

\section*{Acknowledgements}
We would like to thank Til Birnstiel and an anonymous reviewer for comments which lead to a much clearer and generally improved manuscript.
Our work was supported by NASA, under award NNX09AB90G and NNX11AE12G from the Origins of Solar Systems and Astrophysics Theory programs, and by the NSF under award AST-0807471.

     \appendix
     \section[Evolving Disc Temperature]{Calculating the Evolving Disc-temperature Distribution}
     \label{AX_DiscTemp}

The model-disc temperature is calculated using a power-law parameterization of the energy-balanced temperature at the disc midplane.  To do this, the disc is first evolved once using an iterative solver to calculate the energy-balanced temperature at each grid point and each time-step due to passive and viscous heating and radiative cooling.  We then make a power-law fit from the outputted temperature history.

For the energy-balanced temperature, we calculate three components to the disc heating.  The flux of energy due to viscous dissipation near the disc midplane is given by
  \begin{equation}
    F_\mathrm{visc}
  =
    R^2 \nu \Sigma_g \left(\frac{\partial \Omega_g}{\partial R}\right)^2
  \,,
  \label{AX_Fvisc}
  \end{equation}
where the viscosity, $\nu \propto T$, is taken from the alpha-parameterization described in \S\ref{ModelDisc}.  The flux due to heating via stellar illumination is
  \begin{equation}
    F_\mathrm{star}
  =
    2 \sigma_\mathrm{B} \epsilon T_\mathrm{irr}^4
  =
    \frac{L_\star \phi}{2 \pi R^2}
  \,,
  \label{AX_Fstar}
  \end{equation}
where $\sigma_\mathrm{B}$ is the Stefan-Boltzman constant, $T_\mathrm{irr}$ is the passive temperature due to stellar radiation, and the radiative efficiency is assumed to be $\epsilon=1$.  For our disc evolution models we use $L_\star=5 L_\mathrm{solar}$ and set the incidence angle to the disc at $\phi=0.05$ everywhere.  Finally, heating due to background radiation is given by
  \begin{equation}
    F_\mathrm{cloud}
  =
    2 \sigma_\mathrm{B} \epsilon T_\mathrm{cloud}^4
  \label{AX_Fcloud}
  \,,
  \end{equation}
where $T_\mathrm{cloud}=10$K is the background temperature of the cloud environment.

These are used to solve for the midplane temperature of the disc following the approximation derived by \cite{NakamotoNakagawa1994}:
  \begin{eqnarray}
    \sigma_\mathrm{B} T_\mathrm{mid}^4
  &=&
    \frac{1}{2}
    R^2\nu\Sigma_g \left(\frac{\partial \Omega_g}{\partial R}\right)^2
    \left(\frac{3}{8}\tau_R + \frac{1}{2\tau_P}\right)
  \nonumber \\
  &&+ \,
    \sigma_\mathrm{B}\left(T_\mathrm{irr}^4 + T_\mathrm{cloud}^4\right)
  \,,
  \label{AX_midplanetemp}
  \end{eqnarray}
where $T_\mathrm{mid}$ is the temperature at the disc midplane, $\tau_R=\kappa_R\Sigma_g$ is the Rosseland mean optical depth, and $\tau_P$ is the Planck mean optical depth.  For the opacity, $\kappa_R$, we use the piecewise parameterization provided by \cite{BellLin1994} in which $\kappa_R$ depends on both temperature and the mean gas density (for which we use the density at the disc midplane, $\rho_{g,\mathrm{mid}}$).  For simplicity we assume $\tau_P/\tau_R=1$, but note that the specific ratio chosen has negligible impact on the energy-balanced temperature calculated.  The iterative solver seeks a convergence in $T_\mathrm{mid}$ according to Equation~(\ref{AX_midplanetemp}), updating the corresponding values for the midplane density (for a vertically isothermal disc profile), opacity, and disc viscosity at each pass.

The primary focus of this paper is the non-uniform nature of the dust/gas distribution in an evolving disc, an effect which is disregarded for the purposes of the energy-balanced disc-temperature calculation.  In this model of disc temperature, the effects of the dust distribution are felt only in the amount of energy deposited via disc accretion.  If growth of the dust grains were taken into account, dust opacity should drop for larger grain sizes, thereby allowing for less disc heating via accretion.  However, at the time-scales / masses and accretion rates considered in our models, the evolution of disc surface-density contours is reasonably convergent using either a static 
(cooler, 
passive disc) or our evolving--energy-balanced temperature distribution.  If, on the other hand, massive infall of the dust population led to significant dust enhancement in the inner disc, dust opacities in the inner disc should increase.  However, at early times, such infall is comprised solely of large dust grains and so we expect that the effect should be negligible.  Infall of the small-grain population occurs only when the disc is too thin for accretional heating to be a significant contributor to the disc temperature.

Next, we fit the temperature history of a given disc model to a parameterization following the form
  \begin{equation}
    T\left(R,t\right) 
  = 
    T_{AU}\left(t\right)\left(\frac{R}{1\mathrm{AU}}\right)^{q_T\left(t\right)}
  \,.
  \label{AX_powerlawTemp}
  \end{equation}
First $T_\mathrm{AU}$ and $q_T$ are calculated at each time-step using the intersection at $R=0.2$ and 200~AU with the outputted, energy-balanced temperatures.  Next, a quasi-exponential decay in time is fit separately to both $T_\mathrm{AU}\left(t\right)$ and $q_T\left(t\right)$ following the form
  \begin{equation}
    x\left(t\right)
  =
    \left(x_0-x_\infty\right)\exp \left[-\left(\frac{t}{t_x}\right)^{b_x}\right]
  + x_\infty
  \,,
  \label{AX_TPRtimefitting}
  \end{equation}
with $x_0$ and $x_\infty$ representative of the initial and final states of the temperature evolution, and $t_x$ and $b_x$ fitting constants chosen roughly by inspection.  The parameterization constants used for the temperature evolution of each of the disc models of this paper are given below in Table~\ref{AX_tableTPR}.
\begin{table*}
   \begin{center}
   \begin{tabular}{|c|c|c||c|c|c|c|}
   \hline
   &&&&&&
\\
     $M_{D,0}$ ($M_\odot$)
   & $R_d$ (AU)
   & $\alpha$
   & $T_{AU,0}$ (K)
   & $T_{AU,\infty}$ (K)
   & $t_{T_{AU}}$ (yr)
   & $b_{T_{AU}}$
\\
   &&
   & $q_{T,0}$
   & $q_{T,\infty}$
   & $t_{q_T}$ (yr)
   & $b_{q_T}$
\\
   &&&&&&
\\ \hline \hline
     0.01
   & 20
   & $10^{-2}$
   & 458
   & 280
   & $1.9\times 10^5$
   & 1.03
\\
   &&
   & -0.59
   & -0.5
   & $2.2\times 10^5$
   & 1.09
\\ \hline
     0.03
   & 5
   & $10^{-2}$
   & 500
   & 280
   & $2.4\times 10^5$
   & 0.966
\\
   &&
   & -0.61
   & -0.5
   & $2.7\times 10^5$
   & 1.01
\\ \hline
     0.03
   & 20
   & $2\times 10^{-2}$
   & 500
   & 280
   & $2.3\times 10^5$
   & 0.917
\\
   &&
   & -0.61
   & -0.5
   & $2.6\times 10^5$
   & 0.959
\\ \hline
     {\bf 0.03}
   & {\bf 20}
   & {\bf 10$^{-2}$}
   & {\bf 500}
   & {\bf 280}
   & {\bf 3.5$\times$10$^5$}
   & {\bf 0.925}
\\
   &&
   & {\bf -0.61}
   & {\bf -0.5}
   & {\bf 4.0$\times$10$^5$}
   & {\bf 0.966}
\\ \hline
     0.03
   & 20
   & $5\times 10^{-3}$
   & 500
   & 280
   & $5.1\times 10^5$
   & 0.907
\\
   &&
   & -0.61
   & -0.5
   & $5.8\times 10^5$
   & 0.943
\\ \hline
     0.03
   & 40
   & $10^{-2}$
   & 472
   & 280
   & $5.1\times 10^5$
   & 1.01
\\
   &&
   & -0.6
   & -0.5
   & $5.9\times 10^5$
   & 1.07
\\ \hline
     0.09
   & 20
   & $10^{-2}$
   & 600
   & 280
   & $7.3\times 10^5$
   & 1.00
\\
   &&
   & -0.65
   & -0.5
   & $8.1\times 10^5$
   & 1.03
\\ \hline
   \end{tabular}
   \end{center}
\caption[Full-list of Power-law Temperature fitting constants]{Disc parameters and power-law temperature fitting constants (Equations~(\ref{AX_powerlawTemp}) \& (\ref{AX_TPRtimefitting})~) for the disc models used in this chapter.  The $t=0$ temperature is lower if there is initially much less disc mass at small AU.  The time-scaling constants, $t_{T_{AU}}$ and $t_{q_T}$ give a sense of the relative lifetimes of the various disc models, though the lifetimes are more than an order-of-magnitude longer than either of the scaling-constants.  The parameters of the fiducial model are in bold.}
\label{AX_tableTPR}
\end{table*}
These parameterizations are used to calculate disc temperature (and hence, the surface-density evolution of the disc gas) during each of the dust-transport runs (one per grain size per disc model) used to build our dust/gas-evolution simulations.

     \section[Two Cartoon Grain-growth Models]{Two Cartoon Grain-growth Models}
     \label{AX_GrainGrowth}

\begin{figure}
     \begin{center}
     \includegraphics[width=\columnwidth]{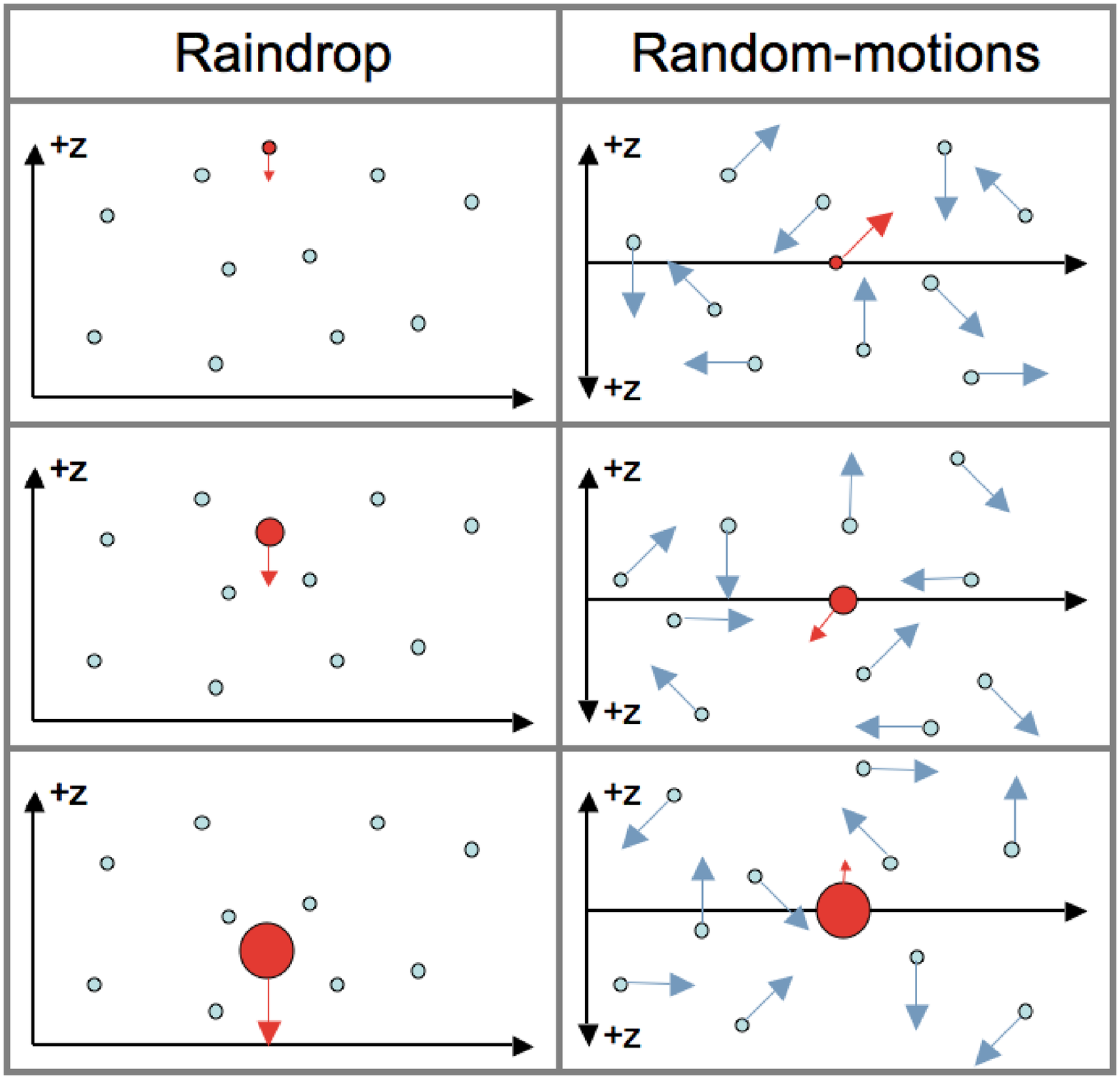}
     \end{center}
     \caption[$\,\,$ Two cartoon grain-growth models]{Depictions of grain growth in two cartoon models.  The modeled, growing grain is highlighted in red, while the background grains from which it grows are shown in blue with black outlines.  Arrows depict the relative magnitude and direction (random in the random-motions model) of grain motions.  Growth in the raindrop model is dependent on grain height, $z$, above the disc midplane, while the grain growing in the random-motions model is assumed to exist always at $z=0$.}
     \label{FAX_graingrowthcartoon}
     \end{figure}

Within the particle-transport simulations, we use two simple-case grain-growth models to place constraints on when and where grains of a given size will appear within the model disc.  These models consider growth of a particle due to differential settling from some height within the disc (called the raindrop model because it mirrors the growth of raindrops falling through the atmosphere), and particle growth due to random, thermal motions at the disc midplane.  The basic schematic for these two models is given in Figure~\ref{FAX_graingrowthcartoon}.

The raindrop model is described in \cite{DullemondDominik2005}.  It considers growth under the extreme condition that only the particle of interest is settling toward the disc midplane and that all other particles remain small and suspended within the disc gas.  In this limit
   \begin{eqnarray}
     \frac{dm_1}{dt}
   &=&
     Z \rho_g\left(z\right)\left|v_\mathrm{settle}\right|\sigma_c
     \,,
   \label{EAX_GGraindrop}
   \\ \nonumber
     v_\mathrm{settle}
   &=&
     -\frac{3\Omega_K^2 z}{4 c_s \rho_g\left(z\right)}\frac{m_1}{\sigma_c}
     \,,
   \end{eqnarray}
where $m_1$ is the mass of the particle, $Z$ is the assumed dust-to-gas mass ratio within the disc, set to $Z=0.01$ for all runs of our grain-growth models, $\rho_g\left(z\right)$ is the gas density local to height $z$ above the midplane, $c_s$ is the sound speed of the gas, and $\sigma_c$ is the collisional cross-section of the growing particle, in this model set to $\sigma_c=\pi s_{d1}^2$, where $s_{d1}$ is the particle radius.  The settling velocity of the grain, $v_\mathrm{settle}$, is defined for the case of Epstein drag.  While the final, $z=0$ size of a grain in this model depends upon the height it is initiated, growth saturates for $z_0\sim 4 H_g$.  In our simulations, we therefore initiate grains from this height for the raindrop-growth calculation.

For the random-motions growth model, we retain the assumption that all background grains remain small ($s_{d2}=0.1 \mu$m), allowing only the particle-of-interest to grow.  Growth in this regime is controlled by the thermal velocity dispersion, $\Delta v_\mathrm{B}$, between grains, and is prescribed by
   \begin{eqnarray}
     \frac{dm_1}{dt}
   &=&
     Z \rho_\mathrm{g,mid} \Delta v_\mathrm{B} \sigma_c \,,
   \label{EAX_GGbrownian}
   \\ \nonumber
     \Delta v_\mathrm{B}
   &=&
     \sqrt{\frac{8 k_\mathrm{B} T \left(m_1+m_2\right)}{\pi m_1 m_2}} \,,
   \end{eqnarray}
where $m_2$ is the mass of a background particle, and the growth model is run at the disc midplane where densities are highest and growth rates will be fastest.  For this model we use the formal definition for the collisional cross-section: $\sigma_c=\pi\left(s_{d1}+s_{d2}\right)^2$.

Note that formally, growth via thermal velocity dispersion is referred to as Brownian grain growth.  However, in growth via Brownian motion, the entire small-grain population grows in size concurrently, so that growth is initially fast, but slows quickly as the thermal velocity drops respective to the larger grain sizes.  However, in turbulent models of grain-growth, growth via random motions can remain fast as the turbulent velocity dispersion increases for the larger size grains.  See, e.g., the static-disc test cases presented in the appendix of \cite{BirnstielDullemondBrauer2010}.  By fixing the background grain population to small sizes in our random-motions grain-growth model, we retain some of the faster grain-growth experienced by larger grains in a turbulent environment, albeit, not to the extent that it occurs in formal turbulent-growth simulations and independent of the specific level of turbulence used within our disc-evolution simulations.

\begin{figure}
     \includegraphics[width=\columnwidth]{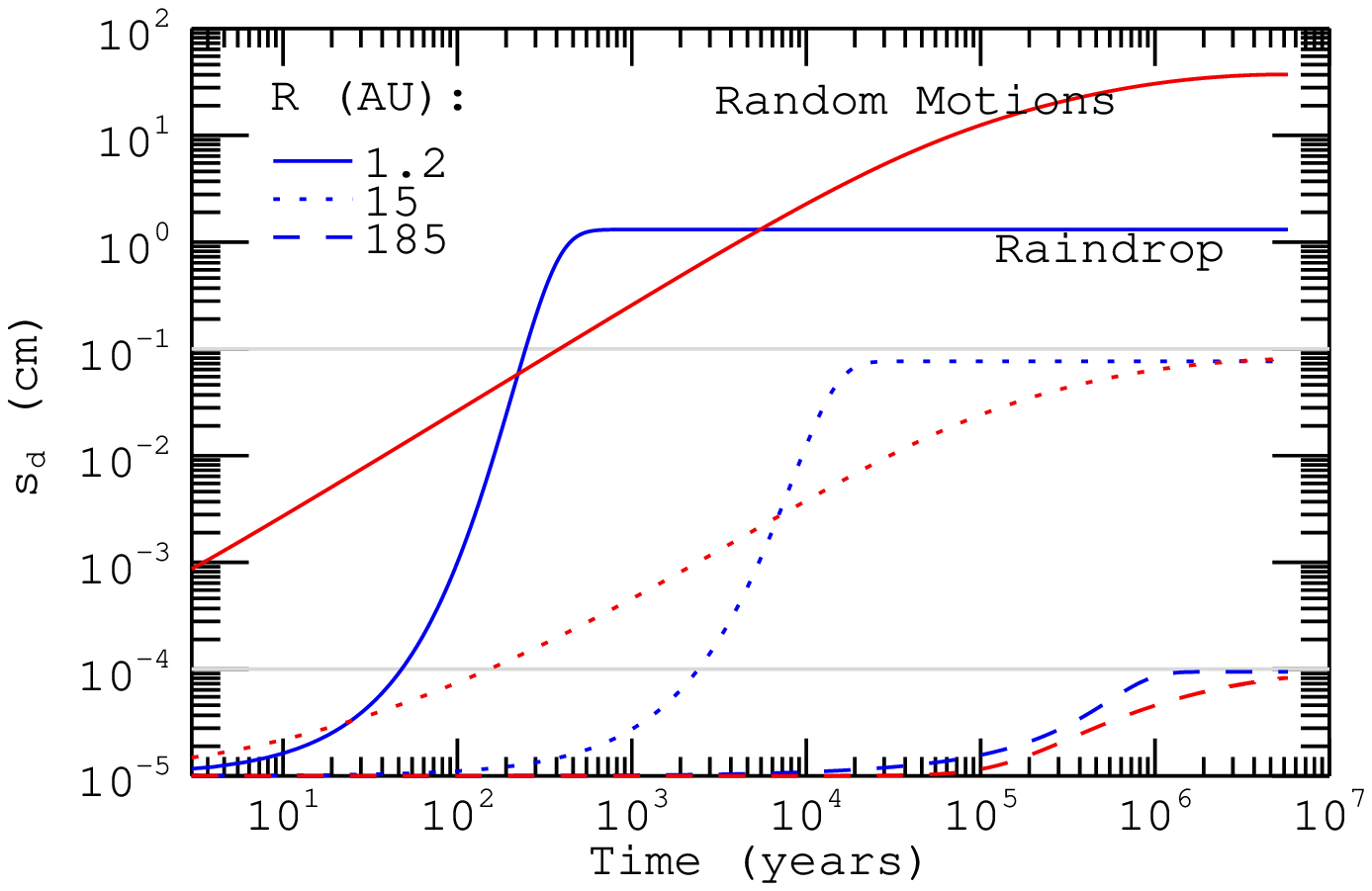}
     \caption[$\,\,$ The cartoon grain-growth models: fiducial]{Grain growth using the two cartoon-grain-growth models.  Blue: the raindrop growth model.  Red: the random-motions growth model.  $\rho_\mathrm{d}=1$ g cm$^{-2}$.  Growth here is modeled within the evolving fiducial disc model.}
     \label{FAX_graingrowthFiducial}
     \end{figure}

In Figure~\ref{FAX_graingrowthFiducial}, 
we plot raindrop growth and random-motions growth within our fiducial disc model at three disc radii.  Because the inner regions of a disc are hotter and denser, particles there grow fastest and to the largest possible sizes.  Particles of a given size are initiated within the particle-transport simulations as soon as either of these grain-growth models grows a particle to that size.

\end{document}